\documentclass[
aps,
pre,
amsmath,amssymb,
reprint,
showkeys
]{revtex4-2}

\usepackage[T1]{fontenc}
\usepackage[utf8]{inputenc}
\usepackage{amsmath}
\usepackage{amssymb}
\usepackage{graphicx}
\usepackage{refstyle}
\usepackage{multirow}
\usepackage[
citecolor=blue,
colorlinks,
linkcolor=blue,
urlcolor=blue,
]{hyperref}
\usepackage{dcolumn}
\usepackage{bm}
\usepackage{soul}
\usepackage[T1]{fontenc}
\usepackage{pifont}
\usepackage[normalem]{ulem} 

\usepackage[table,xcdraw]{xcolor}

\usepackage{array}
\usepackage{mathtools}
\usepackage{lipsum}
\usepackage{tikz}
\usepackage{wrapfig}
\usepackage[T1]{fontenc}
\usepackage[utf8]{inputenc}
\usepackage{amsmath}
\usepackage{makecell}
\usepackage{amssymb}
\usepackage{graphicx}
\usepackage{refstyle}
\usepackage{multirow}
\usepackage{blkarray,booktabs,bigstrut}
\usepackage{placeins}
\usepackage{soul}
\usepackage{mathrsfs}
\usepackage[
citecolor=blue,
colorlinks,
linkcolor=blue,
urlcolor=blue,
]{hyperref}

\usepackage{dcolumn}
\usepackage{bm}
\usepackage{tabularx,booktabs}
\usepackage{multirow}
\usepackage{float}

\graphicspath{{Figures/}}

\begin{document}
	
	
	\title{Prejudice driven spite: A {\color{black}discontinuous} phase transition in ultimatum game }

	\author{Arunava Patra}
    \email{arunava20@iitk.ac.in}
	\affiliation{
		  Department of Physics,
		  Indian Institute of Technology Kanpur,
		  Uttar Pradesh, PIN: 208016,
		  India
		}
	\author{C. F. Sagar Zephania}
	\email{sagarzephania@gmail.com}
	\affiliation{
		  Department of Physics,
		  Indian Institute of Technology Kanpur,
		  Uttar Pradesh, PIN: 208016,
		  India
		}

	\author{Sagar Chakraborty}
	\email{sagarc@iitk.ac.in}
	\affiliation{
		  Department of Physics,
		  Indian Institute of Technology Kanpur,
		  Uttar Pradesh, PIN: 208016,
		  India
		}

	\begin{abstract}
In a mix of prejudiced and unprejudiced individuals engaged in strategic interactions, the individual intensity of prejudice is expected to have effect on overall level of societal prejudice. High level of prejudice should lead to discrimination that may manifest as unfairness and, perhaps, even spite. In this paper, we investigate this idea in the classical paradigm of the ultimatum game which we theoretically modify to introduce prejudice at the level of players, terming its intensity as prejudicity. The stochastic evolutionary game dynamics, in the regime of replication-selection, reveals the emergence of spiteful behaviour as a dominant behaviour via a first order phase transition---a discontinuous jump in the frequency of spiteful individuals at a threshold value of prejudicity. The phase transition is quite robust and becomes progressively conspicuous in the limit of large population size where deterministic evolutionary game dynamics, viz., replicator dynamics, approximates the system closely. The emergence of spite driven by prejudice is also found to persist when one considers long-term evolutionary dynamics in the mutation-selection dominated regime.
	\end{abstract}
\maketitle
\section{Introduction}
Prejudice towards individuals or groups is a kind of preconceived bias that is not entirely based on reason, evidence, or actual experience~\cite{allport1954nature}. These prejudices---whether consciously held explicit prejudice~\cite{Lee2022PSPB} or hardwired implicit prejudice~\cite{cooper1981PB, Dion1972JPSP, Greenwald1995PR}---sometimes go beyond being morally wrong: They are also counterproductive. Moreover, such biases can breed resentment and division. This phenomenon manifests in various forms, including sexism, ageism, and discrimination based on sexual orientation 
\cite{Rudman2012JESP, Nelson2005JSI, Herek2007JSI},
as well as ethnic, racial, nationalistic, and religious extremism 
\cite{Bar2009IUP,Bobo2004PP}.

Prejudice is often understood as a combination of thoughts, feelings, and behaviours directed toward individuals based solely on their group membership. At its core,  prejudice adds an emotional dimension, turning neutral or mildly biased stereotypes into judgments laden with positive or negative feelings.  
Numerous articles and reviews provide evidence supporting the existence of emotions in animals \cite{BEKOFF2000BS, Dsir2002BP, Briefer2012JZ}.
In animals, emotions play a critical role in shaping social behaviours, from fear and aggression to empathy and bonding. Certain behaviours in animals resemble what humans might interpret as prejudice. For example, many species display in-group favouritism, where members of their group are treated more positively than outsiders~\cite{Suarez1999BI, Massen2010PO, Campbell2011PO}. Studies on primates, such as chimpanzees~\cite{Campbell2011PO}, have shown that they are more likely to cooperate with familiar individuals and exhibit aggression toward outsiders.
Animals may not have the ability to form complex stereotypes like humans, but their behaviours, driven by emotions, suggest that at least prejudice might begin as an emotional reaction to group dynamics rather than as a purely mental or social concept. This similarity between humans and animals shows that prejudice may not be unique to humans. Instead, it could be a deeply rooted, emotion-based trait that evolved to help manage complex social situations. 

Prejudice arises through a progression of interconnected social and psychological processes. It begins with stereotypes, which are generalized beliefs about individuals based solely on their group membership. These stereotypes provide a biased framework for evaluating others, often without considering their individuality. Building on this, prejudice adds an emotional layer, turning these stereotypes into judgments filled with positive or negative feelings. For example, people may feel more favourably toward members of their \textit{in-group}—those who share the same social group—while viewing members of an \textit{out-group} less favourably, purely because of their group affiliation. These judgments can occur consciously, where individuals are aware of their bias, or unconsciously, where the implicit bias influences their emotions and decisions without realization. Over time, these prejudiced attitudes may lead to discrimination, where biased thoughts and emotions manifest as actions that advantage or disadvantage individuals based solely on their group membership. This chain of thought, feeling, and action shows how prejudice emerges and perpetuates inequality in society~\cite{Worthy2020GCC}.

Some evidence~\cite{Fehr2002, Gardner2006} suggests that discrimination may contribute to the development of spiteful behaviour. Spite, defined as a behaviour where individuals harm others even at a personal cost. For instance, spiteful behaviours have been observed to evolve under conditions of negative relatedness, where harming others benefits individuals who are less closely related to those they harm \cite{Hamilton1970Nature,Hamilton1996Ox}. Negative relatedness, in this context, highlights how group-based biases can create advantages for some while fostering resentment and spite in others. In a way, relatedness can be a form of discrimination, creating group-based biases and behaviours that lead to unfair and spiteful treatment.

Spiteful behaviour has been observed in both humans~\cite{McAuliffe2014BL, Marcus2014PA} and non-human species~\cite{Sylwester2013JNPE, Gardner2006, Foster2000TREE}. Examples include the red fire ant~\cite{Keller1998}, the parasitoid wasp \emph{Copidosoma floridanum}~\cite{Gardner2006}, and certain bacteria~\cite{Gardner2004}. Over the past few decades, numerous studies~\cite{Smead2019OP, Forber2016PS, Lehmann2006JEB, Kurokawa2023JTB, KNOWLTON1979} have provided valuable insights into the crucial factors that allow spiteful behaviour to persist in evolving populations. {These factors include dynamic networks~\cite{Fulker2021NC}, negative assortment~\cite{Forber2014PRSB}, finite size population~\cite{Smead2012Evolution}, and indirect reciprocity~\cite{Johnstone2004PRSL}, among others.}

As discrimination appears to contribute to spite, and prejudice leads to discrimination, it is reasonable to envisage prejudice as another factor that may bring forth spiteful actions. How can we better understand the relationship between prejudiced attitudes and the emergence of spiteful behaviours? One potential approach of studying this connection is through the ultimatum game (UG)~\cite{Nowak2000Science, Yang2015EL, Henrich2006Science,Han2017PNAS}, a psychological experimental technique commonly used in behavioural economics. The game operates with very simple rules: Two players must decide how to divide a sum of money. One player, the proposer, makes an offer on how to split the amount. If the other player, the responder, accepts the offer, the deal is finalized, and both players receive their agreed shares. However, if the responder rejects the offer, neither player receives anything, and the game ends. Logically, a rational responder should accept any positive amount, no matter how small, since rejecting it results in receiving nothing. Therefore, proposers could, in theory, claim nearly the entire sum for themselves.

Experimental results, however, often contradict rational expectations, as participants frequently make fair offers and reject unfair ones~\cite{Gth1982JEBO,Roth1991AER,Oosterbeek2004EE}. The concept of fairness is not exclusive to humans; evidence from various studies suggests that non-human animals also exhibit behaviours indicative of fairness and equality. For example, brown capuchin monkeys~\cite{Brosnan2003Nature} and chimpanzees~\cite{Brosnan2005PRSB} have been observed to refuse participation in tasks when rewards are distributed unfairly, often displaying frustration when treated less favorably than their peers. Similarly, domestic dogs~\cite{Range2009PNAS} may stop cooperating or obeying commands when they see other dogs receiving better rewards for performing the same task, highlighting their sensitivity to unequal treatment. Despite its simplicity and the predicted rational solution, various models suggest that fairness can emerge as an evolutionary outcome in the UG~\cite{Nowak2000Science, Yang2015EL, Henrich2006Science, Han2017PNAS, Szolnoki_2012_EPL, Szolnoki_2012_PRL,Wang2014,Zhang_2023_CSF}. Some notable mechanisms that contribute to this emergence include defense mechanisms~\cite{Szolnoki_2012_EPL}, reputation mechanisms~\cite{Wang2014}, random resource allocation~\cite{Zhang_2023_CSF}, partner choice and role assignment~\cite{Yang2015EL}.

It is natural to envisage that when group biases are involved, proposers and responders may treat people from the out-group less favorably and give more favourable treatment to those in the in-group. Therefore, the UG can be helpful in understanding how prejudices contribute to discriminatory behaviour, specifically spite. Within the framework of the UG, the existing literature demonstrates that negative assortment promotes fairness, which evolves through the emergence of spiteful behaviour~\cite{Forber2014PRSB}. In other words, when a player is more likely to interact with individuals who are not her type, spiteful strategies tend to arise. However, as seen in empirical studies, organisms frequently display in-group and out-group biases, which are key components of prejudice. The model with negative assortment does not consider the cognitive bias like prejudice which results discriminatory behaviour when interacting with different types. In fact, one would intuit that even in the absence of any assortment, prejudice might drive spite.

In this paper, the phenomenon of prejudice is introduced in the UG and corresponding evolutionary dynamics is investigated to study how prejudice influences player interactions and outcomes in a population. We aim to explore how the presence of prejudice affects decision-making strategies, acceptance thresholds, and overall cooperation within the game. We seek to understand how prejudice can shape social behaviours, influence fairness perceptions, and ultimately impact group cohesion and cooperation in broader evolutionary contexts. This approach may allows us to capture the complex effects of prejudice on spiteful behaviour.

\section{Model: UG with Prejudice}
In the classical UG, two players engage in a strategic interaction to divide a sum of desirable (say, money), normalized to one unit, initially given to one of the players, known as the proposer. The proposer decides what fraction of the money to offer to the other player, called the responder. The responder then makes a choice: either accept the offer and take the proposed amount, leaving the rest for the proposer; or reject the offer entirely, resulting in neither player receiving anything. Here, we shall be interested in the symmetrized version of the UG, where both roles---proposer and responder---are assumed by each player.

Therefore, mathematically~\cite{Nowak2000Science, Rand2013_PNAS}, the strategies of a players is characterized by two parameters: \(o_1\in[0,1]\) representing the amount player \(1\) offers when acting as the proposer, and \(a_1\in[0,1]\) denoting the minimum amount player \(1\) is willing to accept as the responder. The game's outcome is determined by the interaction of the strategies of two players, \(1\) and \(2\). The  payoff of player \(1\), employing strategy \((o_1, a_1)\) while interacting with player \(2\), who employs strategy \((o_2, a_2)\), is based on four possible scenarios:
\begin{itemize}
	\item {Both players accept offers (\(o_1 \geq a_2\) and \(o_2 \geq a_1\)): }  
	Player \(1\)’s payoff is \(1 - o_1 + o_2\), where \(1 - o_1\) is the remaining amount after making their offer, and \(o_2\) is the amount received from player \(2\)'s offer.
	
	\item {Only player \(2\) accepts offer ( \(o_1 \geq a_2\) but \(o_2 < a_1\)):}  
	Player \(1\)’s payoff in this case is only \(1 - o_1\), the remaining amount after making their offer.
	
	\item {Only player \(1\) accepts offer (\(o_1 < a_2\) and \(o_2 \geq a_1\)): }  
Player \(1\)’s payoff is only \(o_2\), the amount received from player \(2\).
	
	\item {Neither accepts offer (\(o_1 < a_2\) and \(o_2 < a_1\)):}  
	This case trivially results in a payoff of zero for player \(1\).
\end{itemize}
Given that the number of possible strategies are uncountable infinite, we simplify the game such that the proposer or responder chooses between two values: a high amount (\(h \le 0.5\)) and a low amount (\(l\)), with \(0 < l < h\). The highest offer or demand, viz., $h=0.5$, represents an equal split of the sum, while the lowest values approach the smallest fraction of the amount. In what follows, it will be helpful to denote components of a strategy as $L$ and $H$ (and not $l$ and $h$ that are to be reserved for payoffs), respectively,  corresponding to $l$ and $h$ as contribution to the payoffs. In other words, $o_i,a_i\in\{L,H\}\,\forall i$. 

Now let us introduce the attribute of prejudice into the picture: A player is deemed either prejudiced (denoted by $\Theta_p$) or unprejudiced (denoted by $\Theta_u$). While a prejudiced player's action should be based on a particular trait of the opponent, in order to keep the number of parameters to a minimum, we consider `being prejudiced' itself as that particular trait. In other words, a prejudiced player base their action on whether their opponent exhibits similar prejudiced traits or not. Specifically, they make offers or demands that favour opponents of their type while disadvantaging those with different type. In contrast, an unprejudiced player make decisions regardless of their opponent's traits, i.e., their action is not influenced by whether the opponent is prejudiced or not. Specifically, the player of this type makes offers and demands that are independent of their opponent's type, ensuring unbiased interactions. Therefore, effectively there are eight possible strategies of a player, say player $i$: $(\Theta_{\theta_i},o_i,a_i)$ where $\theta_i\in\{p,u\}$ and $o_i,a_i\in\{L,H\}$. 

Now for the payoff matrix calculation, we assume that the extent to which offers and demands are adjusted depends entirely on the intensity of prejudice, which we term as \textit{prejudicity}. The prejudicity is represented by the parameter \(e\). The range of \(e\) is such that takes values between 0 to \(h-l\), indicating the extent of the prejudice that influences the payoffs of the players involved. Let us illustrate how the payoff matrix elements are affected. Suppose a $(\Theta_p, \text{H}, \text{H})$ player meets a $(\Theta_p, \text{L}, \text{H})$ player. Note that both are prejudiced---i.e., they are of same trait and so, being prejudiced, they favour each other. Consequently, in this case, the focal player does not change their offer (as $h$ is assumed to be the highest allowed) but reduces their demand to $h-e$, while the opponent increases their offer from $l$ to $l+e$ and decreases their demand from $h$ to $h-e$. Now, with a critical value of prejudicity, \(e_c \equiv {(h-l)}/{2}\), two cases appear: (i) For $0<e < e_c$, the focal player receives $1-h$, and the opponent receives $h$, and (ii) for $e_c\le e<h-l$, the focal player receives \(1-h+l+e\), and the opponent receives \(1+h-l-e\). On the other hand if the focal player,  $(\Theta_p, \text{H}, \text{H})$, meets a $(\Theta_u, \text{L}, \text{H})$ player, the focal player reduces the offer from $h$ to $h-e$, while the unprejudiced opponent keeps their offer and demand intact at $l$ and $h$, respectively. Thus, both players receive zero regardless of the prejudice level. The payoff elements corresponding to all other interactions  can be calculated similarly; one has to additionally keep in mind that a prejudiced focal player with low demand is not allowed to change their demand to $l-e$ as $l$ is assumed to be the smallest allowed. The explicit form of resulting payoff matrix is presented in Appendix~\ref{sec_app:payoff_matrix}. 

We note that the strategies with the pair \((\text{H}, \text{H})\) should be considered \textit{fair} because the proposer offers a high amount, which the responder accepts, ensuring rather equal division of resources. In contrast, a strategy with \((\text{L}, \text{L})\) is deemed \textit{unfair}, favouring the proposer with a larger share while leaving the responder with less. A strategy with \((\text{H}, \text{L})\) may be termed as \textit{altruistic}, as the proposer offers generously (high) but accepts a minimal return (low), prioritizing the responder's benefit at the proposer's cost. Finally, the strategies with \((\text{L}, \text{H})\) represents \textit{spite}, combining a low offer from the proposer with an unreasonably high demand as a responder, reflecting a lack of intent for benefiting the opponent~\cite{Forber2014PRSB} even while incurring a personal cost. 

Next, using the game framework constructed above, we plan to investigate the evolution of spite in both short-term and long-term regimes. Short-term evolution is typically modelled by a replication-selection process~\cite{Traulsen2006_PRE,Nowak2006Book}, which captures changes in trait-frequencies within a population. During this process, the total number of traits remains constant---it is their frequencies that change. In contrast, long-term evolution is driven by a mutation-selection process~\cite{Fudenberg2006_JET, Imhof2009_PRSB}  where, at each time step, a rare mutation arises in a monomorphic population and it either fixates or goes extinct. If the mutant successfully replaces the resident trait, it becomes the new resident of the new monomorphic population, and this process repeats ad infinitum yielding a temporal sequence of monomorphic populations. Importantly, the transient phase of the mutation-selection process—from the appearance of a mutation to its fixation or extinction—corresponds to a short-term evolution because once the mutant arises, its fate is determined by the replication–selection process. We analyze short-term evolution in Section~\ref{sec: Short-term_evolution}. However, even if spite were to emerge and persist under short-term dynamics, its persistence over the long term is not guaranteed. Therefore, to assess the robustness of spite, we subsequently investigate long-term evolution in Section~\ref{Fudenberg}.

\section{Short-term Evolution}
\label{sec: Short-term_evolution} 
The investigation we are after in this paper is evolution of spite in the presence of prejudice. In the framework of UG discussed above, spite is part of a strategy while intensity of prejudice, the prejudicity ($e$), is a parameter affecting the payoff matrix. Thus, within the paradigm of evolutionary game theory, we should now consider a population of players who can have one of the eight different strategies and we should investigate how the fraction of the different strategies evolve over time under replication-selection.

\subsection{Set-up}
In a well-mixed finite population of size \(N\), let the players randomly adopt any strategy from the strategy set $\mathcal{S}=\{\text{F}_p, \text{A}_p, \text{S}_p,\text{U}_p,\text{F}_u, \text{A}_u, \text{S}_u,\text{U}_u\}$ where we have used abbreviated notations for the strategies: $ \text{F}_p = (\Theta_p, \text{H}, \text{H}), \, \text{A}_p = (\Theta_p, \text{H}, \text{L}), \,	\text{S}_p = (\Theta_p, \text{L}, \text{H}), \, \text{U}_p = (\Theta_p, \text{L}, \text{L}), \,\text{F}_u = (\Theta_u, \text{H}, \text{H}), \, \text{A}_u = (\Theta_u, \text{H}, \text{L}), \, \text{S}_u = (\Theta_u, \text{L}, \text{H}),$ \, and  $\text{U}_u = (\Theta_u, \text{L}, \text{L})$. Further, let $s_i$ denote the $i$-th element of set $\mathcal{S}$ which represents the corresponding strategy. For example, $s_1$ corresponds to the 1st element which denotes $\text{F}_p$ strategy, and $s_2$ corresponds to the 2nd element which denotes $\text{A}_p$ strategy, and so on. We assume that the distribution of strategies among individuals in the population at any point in time is \(n_{s_1}, n_{s_2}, \cdots, n_{s_8}\), where \(n_{s_i}\) represents the number of players using strategy $s_i\in \mathcal{S}$; of course, $\Sigma_{i=1}^8n_{{s_i}}=N$. The players randomly interact with each other and accumulate an expected payoff by interacting  with \(N-1\) other individuals.

The expected fitness of player with strategy $s_i\in \mathcal{S}$ is defined to have fitness,
\begin{equation}
	f^w_{s_i}(n_{s_1}, n_{s_2}, \dots, n_{s_8}) = 1-w+w\cdot\frac{1}{N} \sum_{i=1}^{8} n_{s_j} \pi(s_i, s_j),
\end{equation}
where \(\pi(s_i, s_j)\) is the payoff function between strategies $s_i\in\mathcal{S}$ and $s_j\in\mathcal{S}$,  and thus, $\frac{1}{N} \sum_{i=1}^{8} n_{s_j} \pi(s_i, s_j)$ is expected payoff. Also, \(w\in[0,1]\) is the intensity of selection: Higher values of \(w\) favor strategies with higher payoffs, making the selection process more deterministic. Specifically, as \(w \to 1\), the selection becomes strong, while for \(w \to 0\), the process is dominated by neutral drift.

The evolution of strategies in such a well-mixed finite population is well modelled through the Moran process~\cite{Nowak2006Book}. In this process, a random individual is chosen for reproduction with a probability proportional to her relative fitness, and another individual is chosen randomly to die. Subsequently, the offspring replaces the dead individual, so the population size remains constant.
The relative fitness of a player having strategy $s_i\in \mathcal{S}$ is given by, 
\begin{equation}
	p^{r}_{s_i} = \frac{n_{s_i} f^w_{s_i}}{\sum_{k=1}^8 n_{s_k}f^w_{s_k}}.
\end{equation}
Here, we put the superscript $r$ to denote that this is the probability of a player chosen for {reproduction}. Now, the probability that a player is chosen to die is given by,
\begin{equation}
	p^d_{s_i}=\frac{n_{s_i}}{N},
\end{equation}
where, the superscript $d$ mean {death} of an individual.
These two probabilities govern the stochastic evolutionary process of strategies.

In any evolutionary process, mutations---although rare and small---are inevitable. Thus, the offspring of the selected individual (parent, in this context) may not always be identical to the parent because of mutations---errors in the replication process. Offspring that are not identical copies of the parent are regarded as mutants. To introduce mutations in the Moran process, we allow a random mutant to appear at any point in time with probability $\mu$. Thus, at any time step, following set of events occur: (i) a player having strategy $s_i\in \mathcal{S}$ is chosen for reproduction with a probability $p^r_{s_i}$, (ii) the offspring is either an exact copy of the parent with probability $1-\mu$ or a mutant with probability $\mu$ (the mutant adopts one of remaining seven non-parental strategies picked randomly), and (iii) the offspring replaces an individual (having strategy $s_j\in \mathcal{S}$, say) chosen with probability $p^d_{s_j}$ to die.

For our analysis, we choose a small mutation rate, $\mu = 0.05$, without loss of qualitative generality; this choice is validated by checking the outcomes for various other small mutation rates (see Appendix~\ref{sec_app:finite_population}). We do not allow more than one mutation to occur simultaneously in the population, since mutations are rare.  The introduction of rare mutations imparts a nice mathematical property to the Markov chain governing the Moran process: The Markov chain is now ergodic leading to existence of a unique limiting distribution.

All the numerical codes used to generate the results in the manuscript are available at Github~\footnote{\url{https://github.com/ArunavaHub/PrejudiceSpite.git}}.

\begin{figure*}[hbt]
	\centering
	\includegraphics[scale=0.9]{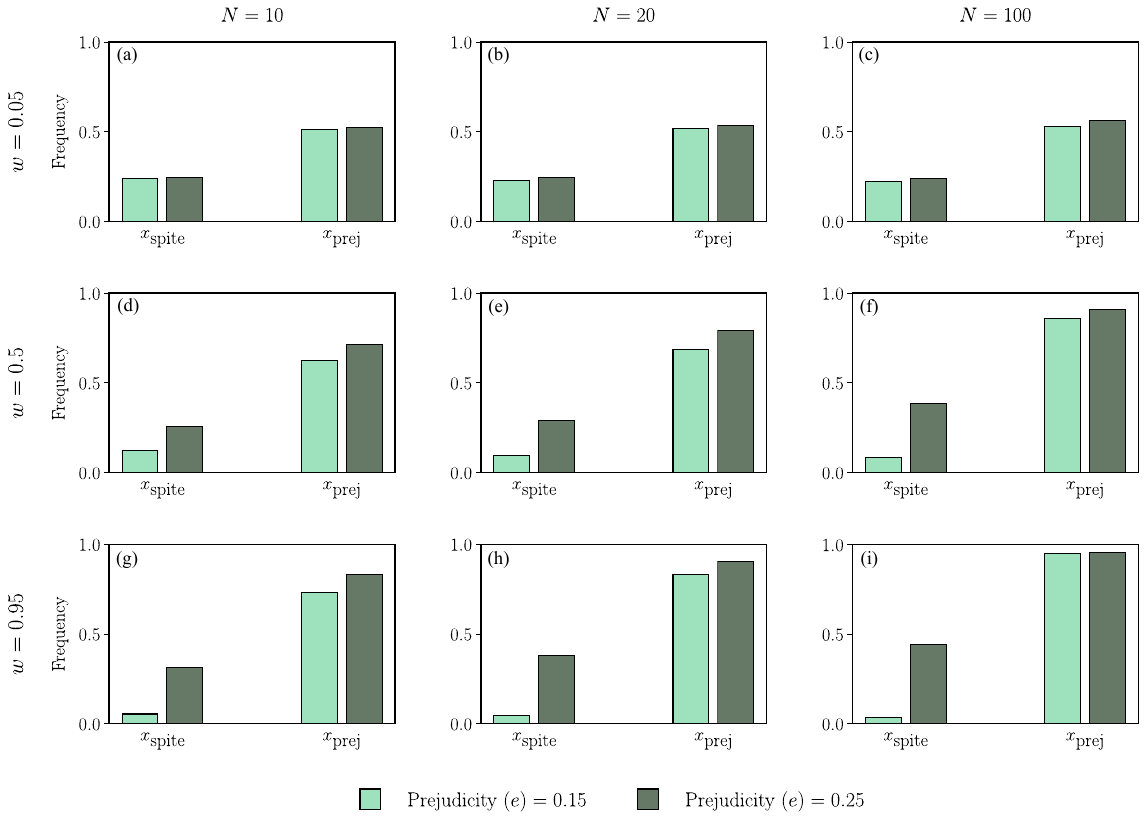}
	\caption {{Histograms illustrating the frequency of spite strategies $(x_{\text{spite}})$ and level of prejudice $(x_{\text{prej}})$ in the short-term evolution of the finite population. We arrange the subplots in $3\times3$ grid corresponding to three selection strengths $w \in \{0.05, 0.5, 0.95\}$ and three population sizes $N\in\{10, 20, 100\}$. The light green and dark green bars, respectively, correspond to prejudicity below $(e = 0.15 < e_c)$ and above $(e = 0.25 > e_c)$ the critical prejudicity ($e_c=0.2$).}}
	\label{fig: Spite_vs_prej} 
\end{figure*}
\begin{figure}[hbt]
	\centering
	\includegraphics[scale=0.32]{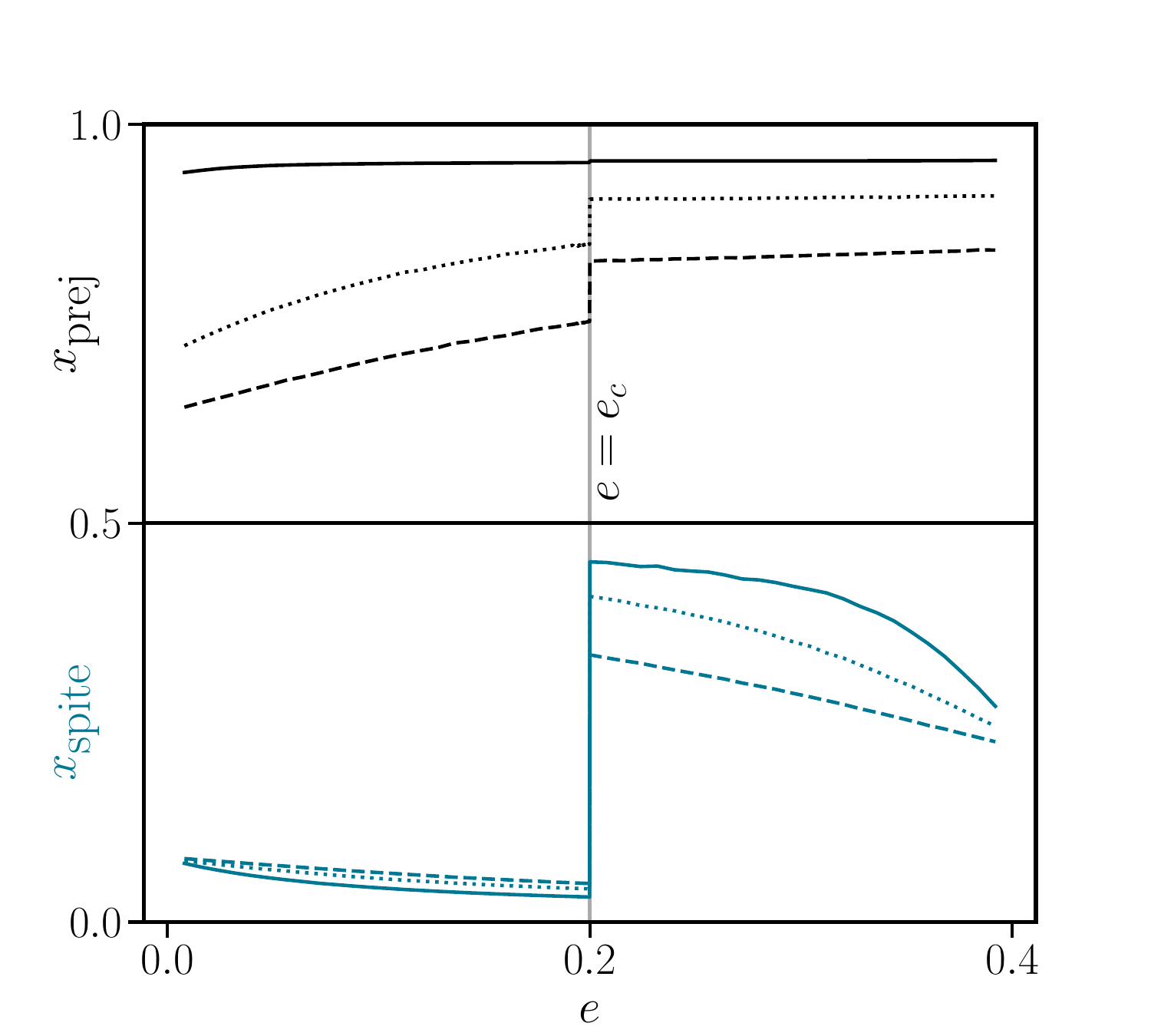}
	\caption {{A {\color{black}discontinuous} phase transition of spite. Plot exhibits in the frequency of spite ($x_{\text{spite}}$) and the level of prejudice ($x_{\text{prej}}$) at the point of critical prejudicity $e=e_c$ for three population size $N\in\{10,20,100\}$. The teal color indicates the frequency of spite, while the black color represents the level of prejudice. Dashed, dotted, and solid lines represent three population sizes: $N=10$, $N=20$, and $N=100$.}}
	\label{fig: Spite_vs_prej_line}
\end{figure}

\subsection{Results}
For the sake of concreteness, without any loss of qualitative generality, we take $l=0.1$ and $h=0.5$ unless otherwise specified. This means that $e_c=0.2$. Our focus is on prevalence of spiteful strategy and level of prejudice in the population; consequently, the mathematical quantities of interest are $x_{\rm spite}\equiv (n_{s_3}+n_{s_7})/N$ and $x_{\rm prej}\equiv(n_{s_1}+n_{s_2}+n_{s_3}+n_{s_4})/N$, respectively, the total frequency of spiteful individuals (both prejudiced and unprejudiced) and the total frequency of prejudiced individuals. 

The numerical simulation of the Moran process, as described above, yields Fig.~\ref{fig: Spite_vs_prej}. The figure presents the frequency distribution of the spiteful strategy and the level of prejudice in a finite population, analyzed across different population sizes (\(N = 10, 20, 100\)) and selection strengths (\(w = 0.05, 0.5, 0.95\)); since there are two payoff matrices corresponding to low $(e < e_c)$ and high  $(e \ge e_c)$ prejudicity,  the figure illustrates two frequency distributions for these cases, specifically for $e =0.15 < e_c=0.2$ (light green bars) and $e = 0.25> e_c=0.2$ (dark green bars). 

For a fixed population size (\(N\)), as selection strength (\(w\)) increases, the frequency of spite rises for high prejudicity and decreases for low prejudicity. Similarly, the level of prejudice increases with selection strength for both low and high prejudicity. Population size (\(N\)) also plays a significant role: For a given selection strength (\(w\)), an increase in population size results in a higher frequency of spite for high prejudicity and a lower frequency for low prejudicity. Meanwhile, the level of prejudice consistently increases with population size across both levels of prejudicity. Overall, the figure highlights the dynamic interplay between selection strength, population size, and prejudicity in shaping the prevalence of spite and prejudice within the population.

As is expected, under weak selection (\(w = 0.05\)), $x_{\rm spite}$ approaches \(0.25\), while $x_{\rm prej}$ is approximately \(0.5\), because in this regime, weak selection is predominantly influenced by random drift, leading to an equal probability for all the eight strategies. In contrast, at higher selection strengths (\(w = 0.95\)) and larger population sizes (\(N = 100\)), it is observed that prejudiced players dominate the population, as indicated by the level of prejudice approaching a value close to one. Also, the frequency of the spiteful strategy stabilizes around 0.5, signifying that approximately half of the population adopts the spiteful strategy. The contrasting scenarios in small and large populations may be qualitatively comprehended as follows. In smaller populations (\( N \) is low), mutations have a stronger impact as they represent a larger fraction of the population, and genetic drift increases the chances of new strategies spreading. This higher mutation-driven introduction of strategies enhances diversity within the population. As diversity grows, dominant strategies like spite face greater competition, reducing their frequency. In larger populations (\( N \) is high), the effect of individual mutations is diluted, and the spread of new strategies relies more on selective advantage. This difference makes the emergence and influence of new strategies more noticeable in small populations, while dominant strategies are more stable in larger populations. 

Above observations naturally motivate us to check the full variation of the frequency of spiteful strategies  and the level of prejudice as functions of prejudicity. Fig.~\ref{fig: Spite_vs_prej_line} showcases the variation in $x_{\rm spite}$ (blue curves) and $x_{\rm prej}$ (black curves) as functions of prejudicity (\(e\)) for three different population sizes (\(N=10\) (dashed), \(N=20\) (dotted), and \(N=100\) (solid)). These curves highlight a critical threshold at \(e = e_c\), where the frequency of spiteful strategies increases sharply. The peak value of frequency of spite grows with increasing \(N\); the jump also becomes sharper with increase in $N$. A noticeable jump in the frequency of prejudice occurs at \(e = e_c\); however, this jump is more pronounced for smaller values of \(N\) and with increasing $N$, the jump vanishes. 
 
It is intriguing that the dominant emergence of spiteful strategies beyond the critical prejudicity \( e_c \) can be understood as a {\color{black}discontinuous} phase transition. At this threshold, the system undergoes a qualitative shift in its evolutionary dynamics: spiteful behaviour, previously negligible, appears suddenly and grows sharply. We see $x_{\rm spite}$ is the order parameter that shows discontinuous jump as a system defining parameter $e$ changes: the sharper and robust jump with increasing $N$ (towards a kind of analogous thermodynamic limit) makes this interpretation even more convincing. It is worth pointing out the dependence of this phase transition on the difference \( (h-l) \) rather than individual values of \( h \) and \( l \) underscores the universality of this transition. In this context, a detailed investigation of the frequency distribution of spiteful strategies is examined for varying the minimum offer or demand ($l$) in Appendix~\ref{sec_app:finite_population} which also presents the frequency distribution of all the strategies across different population sizes (\(N\)) and varying selection strengths (\(w\)).

In passing, we note that after the phase transition, the value of $x_{\rm spite}$ slightly decreases as \(e\) approaches its maximum value (\(h-l\)). At high values of $e$, the prejudiced players become predominant in the population, see the black lines in Fig.~\ref{fig: Spite_vs_prej_line}. Therefore, the frequency of spite strategy is largely driven by the prejudiced players. The prejudiced players increase the offer almost to $h$ and decrease the demand almost to $l$ against a prejudiced opponent as $e$ approaches $(h-l)$. This results in the payoffs for all prejudiced strategies against prejudiced players becoming nearly equivalent to the payoff of the strategy $( \Theta_p, H, L)$. In fact, the payoffs of all  the prejudiced strategies become equal at $e=(h-l)$.  As a result, the fitness difference between all prejudiced strategies decreases as $e$ approaches $(h-l)$. It leads to a rise in the frequency of prejudiced fair and altruistic strategies, while the dominance of prejudiced spite and unfair strategies decreases.  Consequently, we observe a decline in the overall frequency of spite in the population.

\section{Long-term Evolution}\label{Fudenberg}
The robust appearance of spite when prejudicity crosses a threshold value is a far general feature: It can be seen even when we consider mutation-selection dominated long-term stochastic evolutionary process.

Specifically, we are thinking about long-term evolution of the strategies in a well-mixed population of fixed size. Initially, all the individuals in the population adopt an identical strategy, i.e., the population is monomorphic. Suppose a rare mutation in strategy occurs. Here, we consider that no further mutation  occurs until the first mutant either invades the resident population or becomes extinct. If the mutant invades the resident population, then the mutant becomes a new resident in the population. Subsequently, another random mutant comes, which either invades or goes to extinction, and thus the mutation-selection process goes on. Therefore, on a large time scale, the population is always monomorphic, and there is a continuous transitions between the eight monomorphic population states. In the context of present study, one would like to see if high prejudicity leads to emergence of larger probability of monomorphic population with spiteful strategies when the transitions have settles at a steady-state.

One of the well-accepted ways for investigating this conceptual framework is the Imhof--Fudenberg--Nowak process~\cite{Fudenberg2006_JET, Imhof2009_PRSB} which we now set up tuned to our purpose.
\subsection{Set-up}
Consider a well-mixed homogeneous population of $N$ individuals with a particular resident strategy, $s_i$. Suppose a mutant strategy, $s_j\ne s_i$, appears. Then, fixation probability of the mutant $\rho(s_i,s_j)$ is given by~\cite{Nowak2004} the expression
\begin{equation}
\rho(s_i,s_j)=\left(1+\sum_{i=1}^{N-1}\prod_{k=1}^{i} \exp(-\beta[\pi_{s_j}(k)-\pi_{s_i}(k)])\right)^{-1},
\end{equation}%
where
\begin{subequations}
\begin{eqnarray}
	\pi_{s_i}(k)&=&\frac{N-k-1}{N-1}\pi({s_i,s_i})+\frac{k}{N-1}\pi({s_i,s_j}),\\
	\pi_{s_j}(k)&=&\frac{N-k}{N-1}\pi({s_j,s_i})+\frac{k-1}{N-1}\pi({s_j,s_j}).
\end{eqnarray}
\end{subequations}
which, respectively, are the expected payoffs of resident and mutant when the population has $k$ mutants. Here we have assumed pairwise imitation process~\cite{Traulsen2006_PRE} such that the fitnesses of the resident and the mutant are, respectively, given by $\exp(\beta 	\pi_{s_i})$ and $\exp(\beta 	\pi_{s_j})$. The parameter $\beta\in(0,\infty)$ acts as a measure of the selection strength. 

In effect, we have a discrete-time Markov chain with eight states---each corresponding to a monomorphic population with the strategies in $\mathcal{S}$. Let $\bm{\alpha}(0)$ be the probability vector over the states at initial time $t=0$, then $\bm{\alpha}(t)$ at any future time $t$ is given by $\bm{\alpha}(t)=\bm{\alpha}(0){\sf R}^t$ where ${\sf R}$ is the transition matrix. The elements, $r(s_i,s_j)$, of ${\sf R}$ are given as~\cite{Fudenberg2006_JET}
\begin{equation}
	r(s_i,s_j) =
	\begin{cases}
		\mu_{s_is_j}\rho(s_i,s_j)~~~~~~~~~~~~~~~~~~~\text{if}~{s_i}\neq{s_j}, & \\
		1-\sum_{s_j\ne s_i}\mu_{s_is_j}\rho(s_i,s_j),~~~\text{if}~{s_i}= {s_j}; & 
	\end{cases}   
\end{equation}%
where, $\mu_{s_is_j}$ is mutation rate of strategy $s_j$ in the monomorphic population with strategy $s_i$. This Markov chain is evidently ergodic and hence a limiting distribution, $\bm{\alpha}_\infty\equiv\lim_{t\to \infty}\bm{\alpha}(t)$, exists uniquely. Since each element, $\alpha_i$, of $\bm{\alpha}$ measures the frequency of corresponding population state with strategy $s_i\in\mathcal{S}$. In this notation, the average spite level at time $t$ is, thus, the sum $(\alpha_3+\alpha_7)$ of vector elements corresponding to states $s_3$ ($S_p$) and $s_7$ ($S_u$).

\begin{figure}[hbt]
	\centering
	\includegraphics[scale=0.28]{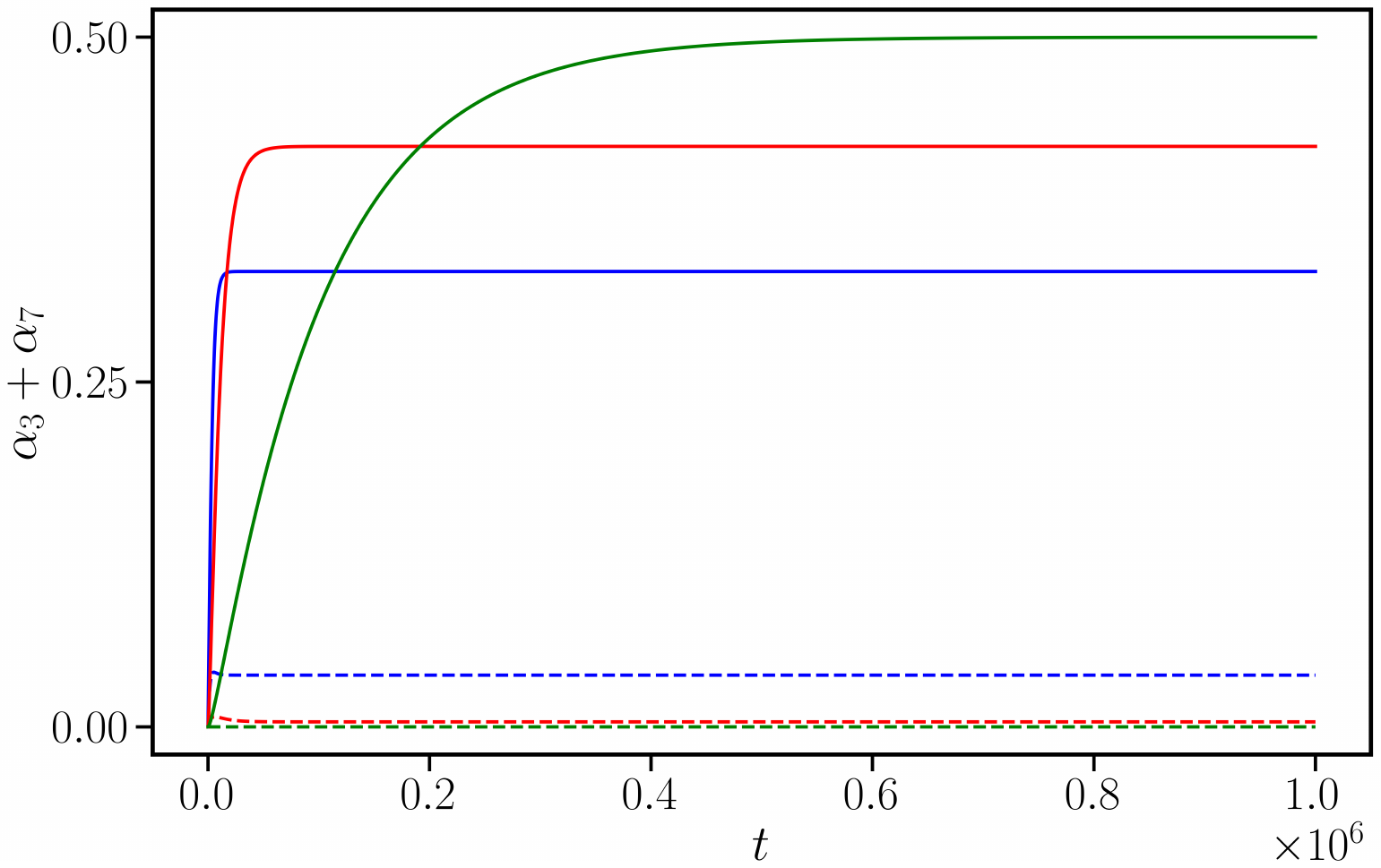}
	\caption {Emergence of spite beyond threshold prejudicity: Plot showcases the frequency of spite strategies over time for three population sizes $N\in\{10,20,100\}$ in the Imhof--Nowak--Fudenberg process. The frequency of spite means how many times the spiteful monomorphic population appears up to time $t$. {Dashed and solid lines, respectively, correspond to the prejudicity $e=0.15<e_c$ and $e=0.25>e_c$.} Blue, red, and green colors represent the population sizes $N=10$, $N=20$, and $N=100$, respectively. We keep the parameter selection strength $\beta=0.95$ and mutation rate $\mu_{s_i s_j} = 10^{-3}$.}
	\label{fig: SFreq_timej}
\end{figure}

\subsection{Results}
It makes sense, in order to understand the long-term evolution of spite driven by prejudice, to assume that all individuals are initially unprejudiced and adopt the fair strategy ($s_5$, i.e., $\text{F}u$). For the emergence of spite, mutation must then occur in the population. In this work, we restrict our analysis to weak selection strengths: $\beta<1$. Therefore, by choosing $s_i=s_5$ in the aforementioned setup, we numerically run the process with $\mu_{s_is_j}=10^{-3}$ and $\beta=0.95$, with no loss of qualitative generality within the working range of selection strength. A comparative analysis of the long-term evolution of spite across higher selection strengths and mutation rates is provided in Appendix~\ref{sec_app:long_term_finite_population}.

In order to gain insight into a comparative perspective on how the dynamics of spiteful strategies vary with population size and prejudicity over time, we generate Fig.~\ref{fig: SFreq_timej}. There we illustrate the probability of the spiteful populations as a function of time for three different population sizes (\(N = 10, 20, 100\)), and two distinct values of prejudicity---low (\(e = 0.15\)) and high (\(e = 0.25\)). We observe that for low prejudicity, the frequency of spiteful strategies remains nearly zero for all values of \(N\). In contrast, for high levels of prejudicity, the frequency converges rapidly, particularly for smaller population sizes. In the asymptotic limit, however, the frequency of spite for \(N=100\) exceeds that for \(N=20\) (compare the green and red solid lines in Fig.~\ref{fig: SFreq_timej}), and likewise, the frequency for \(N=20\) exceeds that for \(N=10\) (compare the red and blue solid lines). This indicates that spiteful behaviour increases as the population size grows, with larger populations leading to a higher prevalence of spiteful strategies. Note that this is true when the selection strength is moderate; further analysis for stronger selection is provided in Appendix~\ref{sec_app:long_term_finite_population}. It has also been found that the maximum frequency of spite stabilizes at 0.5. Since the mutation-selection process is ergodic, the results remain unchanged in the asymptotic limit even if the individuals initially adopt any strategy other than $s_5$.

\begin{figure}[hbt]
	\centering
	\includegraphics[scale=0.3]{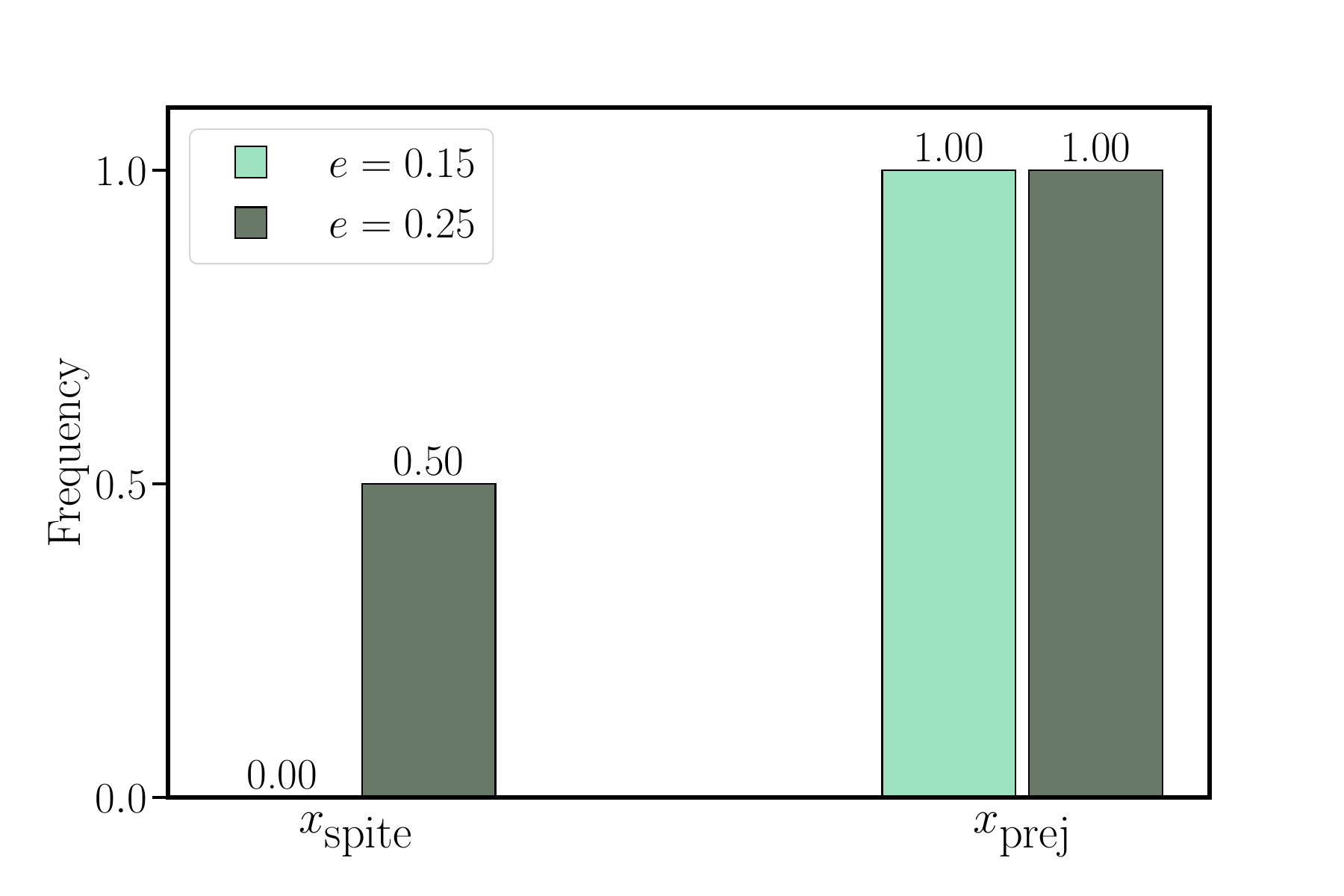}
	\caption {Histogram illustrating the frequency of spiteful strategies $(x_{\text{spite}})$ and level of prejudice $(x_{\text{prej}})$ in the long-term evolution of a finite population. The light green and the dark green bars, respectively, represent the scenarios with prejudicity below $(e = 0.15 < e_c)$ and above $(e = 0.25 > e_c)$ the critical prejudicity ($e_c$). We fix the population size $N=100$, selection strength $\beta=0.95$ and mutation rate $\mu_{s_is_j} = 10^{-3}$.} 
	\label{fig: Spite_vs_prej_IMF} 
\end{figure}

The limiting probability distribution of the spiteful populations and level of prejudice at large times (say, \(t = 10^{6}\)) and  for a population size of \(N = 100\) are depicted in Fig.~\ref{fig: Spite_vs_prej_IMF}. Interestingly, the results closely align with those obtained from the Moran process (see Fig.~\ref{fig: Spite_vs_prej}(i)). Therefore, even in the long-term evolution of the system, the prejudiced spite strategy tends to dominate when the prejudicity is high. (For frequency distribution of other strategies Appendix~\ref{sec_app:finite_population} may be referred.)

\section{Discussion and Conclusions}
We have theoretically examined the role of prejudice in establishing evolutionary robust spite. Technically, we introduce prejudice in a one-shot symmetrized UG at the level of players such that the players discriminate in offering and accepting the offer based on their opponent's observable traits. In one-shot symmetrized UG, each player can offer and demand; and therefore, it has four possible strategies: \emph{fair}, \emph{altruistic}, \emph{spite}, and \emph{unfair}. In real social systems, individuals often exhibit preferential treatment toward in-group members and hostility toward out-group members, driven by simultaneous in-group and out-group biases~\cite{Tajfel1971, Brewer1999}, which are often regarded as the roots of prejudice. In accordance with it, we have assumed that a prejudiced player favours an opponent with the same trait by increasing low offers and reducing high demand. In contrast, the prejudiced player reduces the high offer and raises the demand against any opponent who has different trait. The modifications of offer and demand rely entirely on the player's intensity of prejudice, which we term in this paper as \emph{prejudicity} ($e$). 
	
Subsequently, we consider a well-mixed finite population of prejudiced players with four strategies and unprejudiced players with four strategies and employ the Moran process with rare mutation to study the evolution of frequencies of strategies in the short term. In addition, we study the long-term evolution of the strategies in finite population using the Imhof--Fudenberg--Nowak process~\cite{Fudenberg2006_JET, Imhof2009_PRSB}. In different population sizes, we comprehensively investigate the evolution of the strategies under weak and strong selection strengths. Our model reveals that spite can be sustained in the population due to the prejudice inherent in the individuals. We find a critical value of prejudicity $(e_c)$ above which the spite rises suddenly. In fact, at this critical value of prejudicity, we observe a {\color{black}discontinuous} phase transition in the frequency of spite in the population, and the phase transition becomes prominent in a strong selection regime and is independent of the population size. It is furthermore interesting to note that the critical value is not decided solely by the highest or the lowest offer (or demand) in the population, but rather by the  the difference between the highest and lowest offer (or demand). 	

The mechanism of prejudice adopted in this paper may be referred to as \emph{symmetric prejudice}, since players display both tendencies: favouring in-group members and discriminating against out-group members. An alternative scenario could be that the players maintain their original strategies while interacting with the same type of individuals and adjust their behaviour only when facing non-prejudiced individuals. Such a mechanism can be termed as \emph{(totally) asymmetric prejudice}, since individuals exhibit negative reactions exclusively toward out-group members. Although not presented herein, we had also examined the evolution of spite under totally asymmetric prejudice and observed that no sharp rise in the frequency of spite is observed. In summary, totally asymmetric prejudice does not promote the emergence of spite, whereas symmetric prejudice is necessary to trigger a sharp increase in spite frequency beyond a critical prejudicity. To put it other words, both in-group favouritism and out-group hostility may be essential for the evolution of spite. Of course, a more detailed technical presentation of such comparison would be a useful future endeavour.
	
Under conditions of large population size and strong selection, the influence of stochastic events, such as mutations or random sampling, is tamed. Consequently, the evolutionary dynamics of the finite population should resemble those of an infinite population governed by replicator dynamics, where frequencies of strategies evolve purely based on their relative payoffs without significant influence from random fluctuations. Thus, both the Moran process in finite populations and the replicator dynamics in infinite populations are expected to yield comparable evolutionary outcomes under these conditions. Furthermore, this similarity facilitates insights about the spite strategy from the viewpoint of its evolutionary stability, whose well-known tight connection~\cite{Hofbauer1979,Zeeman1980} with the dynamic stability of corresponding population state---a fixed point of the replicator equation---is contextually elaborated in Appendix~\ref{sec_app:infinite_population}. Furthermore, it is worth pointing out that in the face of continuous random mutations, evolutionary stability, as encapsulated in the ideas of stochastic stable equilibrium~\cite{Foster1990TPB} and long-run equilibrium~\cite{Kandori1993}, is what is achieved in the Moran process with mutation discussed in this paper.

The fact that the phase transition becomes more sharp and robust as the system size increases is quite reminiscent of phase transitions in statistical mechanics. In the context of phase transitions, the analogy between systems of players under evolutionary dynamics and statistical mechanical systems is a subject of intense current research which has been well-reviewed~\cite{Sazbo2007PR, Sazbo2016PR, Scott2022SChaos} in literature. The deviation from rationality leads to noisy decision-making and hence, this deviation can be viewed as an inherent temperature. Also, the collective payoff of the population can serve as (negative) energy in this language. In the special case of potential games~\cite{Sazbo2016PR}, such correspondences facilitates more close mathematical analogy: One can constructed even a partition function for the game-theoretic system where the change in temperature may lead to phase transitions. The phase transition we have witnessed in this paper is novel in some respect. The well-mixed population playing the UG can be seen as an all-to-all coupled eight-state Potts model~\cite{Sazbo2007PR} at constant temperature. The order parameter, characterizing the degree of organized spite, is the frequency of spiteful strategies and the control parameter is prejudicity. The sudden discontinuous jump in the frequency at critical prejudicity ($e_c$) can, thus, be interpreted as a {\color{black}discontinuous} phase transition. The jump is accompanied by the discontinuous change in the payoff elements, analogically viewed as contributing to the energy of the system. We also note that the phases---characterized by different degrees of spite---are distinct limiting stationary states of respective ergodic stochastic processes; unlike what happens in many non-equilibrium phase transitions~\cite{Scott2022SChaos}, the phase transition here does not involve any absorbing state.
	
Thus, prejudice plays a significant role in shaping social dynamics. Prejudicial attitudes are not confined to humans~\cite{Dovidio2002, Abbink2019, Hagiwara2022}; similar patterns appear in non-human species as well~\cite{Campbell2011PO, vonSchmalensee2025}. For example, chimpanzees exhibit in-group favouritism and out-group hostility, demonstrating empathy toward members of their own group, aggression toward outsiders, and even discriminatory behaviours such as contagious yawning, which differ between in-group and out-group members~\cite{Campbell2011PO}. These tendencies likely stem from deeply rooted cognitive mechanisms associated with in-group and out-group biases, which form the foundation of prejudice. Our findings suggest that when such biases in biological species are interpreted as forms of prejudice—quantified by an intensity parameter that measures the degree of in-group preference and out-group hostility—there exists an interesting critical threshold of this intensity. Beyond this threshold, spiteful behaviour emerges abruptly through a discontinuous phase transition. This implies that in real social systems, whether human or non-human, strong intensity of prejudice can trigger---without any prior perceptible signature or warning---sudden appearance of antisocial behaviour like spite. This phenomenon is reminiscent of the tipping point which is an explosive current research topic of nonlinear dynamics and complex systems~\cite{SCHEFFER2020}, especially in the context of climate and ecosystem changes~\cite{Scheffer2001, Dakos2019, Rocha2018}: Tipping point is a critical threshold at which a system undergoes a sudden, dramatic, and often irreversible shift to a new, contrasting state. Thus, just like in the case of tipping points, our work highlights the importance of understanding and mitigating prejudice within the context of cultural evolution in order to avoid sudden emergence of undesirable spite behaviour.

While our current model assumes a well-mixed population, in reality, interactions are typically structured spatially or more broadly, within networks, as individuals are likely to preferentially interact with certain specific players. Therefore, extending our framework to incorporate network-structured populations~\cite{Szolnoki_2012_PRL, Szolnoki_2012_EPL} would reveal the role of network reciprocity in the emergence of spite. For regular networks in the infinite population limit, this effect can be analyzed easily since the payoff matrix merely transforms rendering its entries to become functions of the degree of network~\cite{Ohtsuki2006}. More complex network structure, however, would likely require extensive numerical investigation. The long-term evolutionary dynamics in such system requires calculation of fixation probabilities of mutants on graphs~\cite{Lieberman2005}; the probabilities may depend on the exact update rule of the short-term evolutionary dynamics. Additionally, our model assumes that player interactions are one-shot, with players employing memory-less strategies since their actions are not influenced by past moves. In contrast, real-world scenarios often involve repeated interactions with the same opponents, allowing players to use past experiences to inform future decisions. Introducing repeated interactions and memory-based strategies into our model---possibly in network-structured population---could provide valuable insights into how direct reciprocity influences the co-evolution of prejudice~\cite{Chadefaux2012} and spite.

\acknowledgements
Authors thank Supratim Sengupta for insightful discussions. AP thanks CSIR (India) for the financial support in the form of Senior Research Fellowship and SC acknowledges the support from SERB (DST, govt. of India) through project no. MTR/2021/000119. 
\appendix
\section{8$\times$8 Payoff Matrices}
\label{sec_app:payoff_matrix}
In the two-player symmetrized ultimatum game (UG) with prejudice and with the players restricted to have only two discrete actions---high ($h$) or low ($l$)---the explicit forms of the payoff matrix are given in Table~\ref{tab_1}. Here, for brevity the strategies are denoted as follows: $ \text{F}_p = (\Theta_p, \text{H}, \text{H}), \, \text{A}_p = (\Theta_p, \text{H}, \text{L}), \, \text{S}_p = (\Theta_p, \text{L}, \text{H}), \, \text{U}_p = (\Theta_p, \text{L}, \text{L}), \,\text{F}_u = (\Theta_u, \text{H}, \text{H}), \, \text{A}_u = (\Theta_u, \text{H}, \text{L}), \,	\text{S}_u = (\Theta_u, \text{L}, \text{H}),$ \, and  $\text{U}_u = (\Theta_u, \text{L}, \text{L})$. 
\begin{table*}
	\begin{center}
		\begin{blockarray}{ccccccccc}	
			&$F_p$ & $A_p$  & $S_p$ & $U_p$ & $F_u$ & $A_u$ & $S_u$ & $U_u$ \\
			\begin{block}{c[cccccccc]}
				$F_p $&$1$ & $1$ & $1 - h$ & $1 - h$ & $h$ & $1 + e$ & $0$ & $1 - h+e$\\
				&\\
				$A_p$&$1$ & $1$ & $1  - h + l+ e$ & $1  - h + l+ e$ & $h$ & $1 + e$ & $0$ & $1 - h+e$\\
				&\\
				$S_p$&$h$ & $1  + h - l- e$ & $0$ & $1 - l-e$ & $h$ & $1 + h - l$ & $0$ & $1 - l$\\
				&\\
				$U_p$&$h$ & $1  + h - l- e$ & $ l+e$ & $1$ & $h$ & $1 + h - l$ & $0$ & $1 - l$\\
				&\\
				$F_u$&$1 - h$ & $1 - h$ & $1 - h$ & $1 - h$ & $1$ & $1$ & $1 - h$ & $1 - h$\\
				&\\
				$A_u$&$1 - e$ & $1 - e$ & $1 - h + l$ & $1 - h + l$ & $1$ & $1$ & $1 - h + l$ & $1 - h + l$\\
				&\\
				$S_u$&$0$ & $0$ & $0$ & $0$ & $h$ & $1 + h - l$ & $0$ & 
				$1 - l$\\
				&\\
				$U_u $&$ h-e$ & $h-e$ & $l$ & $l$ & $h$ & $1 + h - l$ & $l$ & $1$\\
			\end{block}
		\end{blockarray} 
	\end{center}
	
	\begin{center}
		\begin{blockarray}{ccccccccc}	
			&$F_p$ & $A_p$  & $S_p$ & $U_p$ & $F_u$ & $A_u$ & $S_u$ & $U_u$ \\
			\begin{block}{c[cccccccc]}
				$F_p$&$1$ & $1$ & $1  - h + l+ e$ & $1  - h + l+ e$ & $h$ &$ 1 + e$ & $0$ &$ 1 - h + e$\\
				&\\		
				$A_p$&1 & $1$ & $1  - h + l+ e$ & $1  - h + l+ e$ & $h$ & $1 + e$ & $0$ & $1  - h + e$\\
				&\\		
				$S_p$&$1  + h - l- e$ & $1  + h - l- e$ & $1$ & $1$ & $h$ & $1 + h - l$ & $0$ & $1 - l$\\	
				&\\		
				$U_p$&$1  - h + l+ e $& $1  - h + l+ e$ & $1$ & $1$ & $h$ & $1 + h - l$ & $0$ & $1 - l$\\		
				&\\	
				$F_u$&$1 - h$ & $1 - h$ & $1 - h$ & $1 - h$ & $1$ & $1$ & $1 - h$ & $1 - h$\\	
				&\\		
				$A_u$&$1 - e$ &$ 1 - e$ & $1 - h + l$ & $1 - h + l$ & $1$ & $1$ &$ 1 - h + l$ & $1 - h + l$\\	
				&\\		
				$S_u$&$0$ & $0$ & $0$ & $0$ & $h$ & $1 + h - l$ & $0$ &$ 1 - l$ \\
				&\\
				$U_u$&$ h-e$ & $ h-e$ &$ l$ & $l$ &$ h$ & $1 + h - l $& $l$ & $1$\\
			\end{block}
		\end{blockarray} 
		\caption{The upper payoff matrix is for prejudicity $e$ such that  $0< e<e_c\equiv(h-l)/2$ and the lower one is for prejudicity $e$ such that  $e_c\equiv(h-l)/2\le e<(h-l)$.}
		\label{tab_1}
	\end{center}
\end{table*}

\section{Short-term Evolution in Finite Population} \label{sec_app:finite_population}
\begin{figure*}
	\centering
	\includegraphics[scale=0.99]{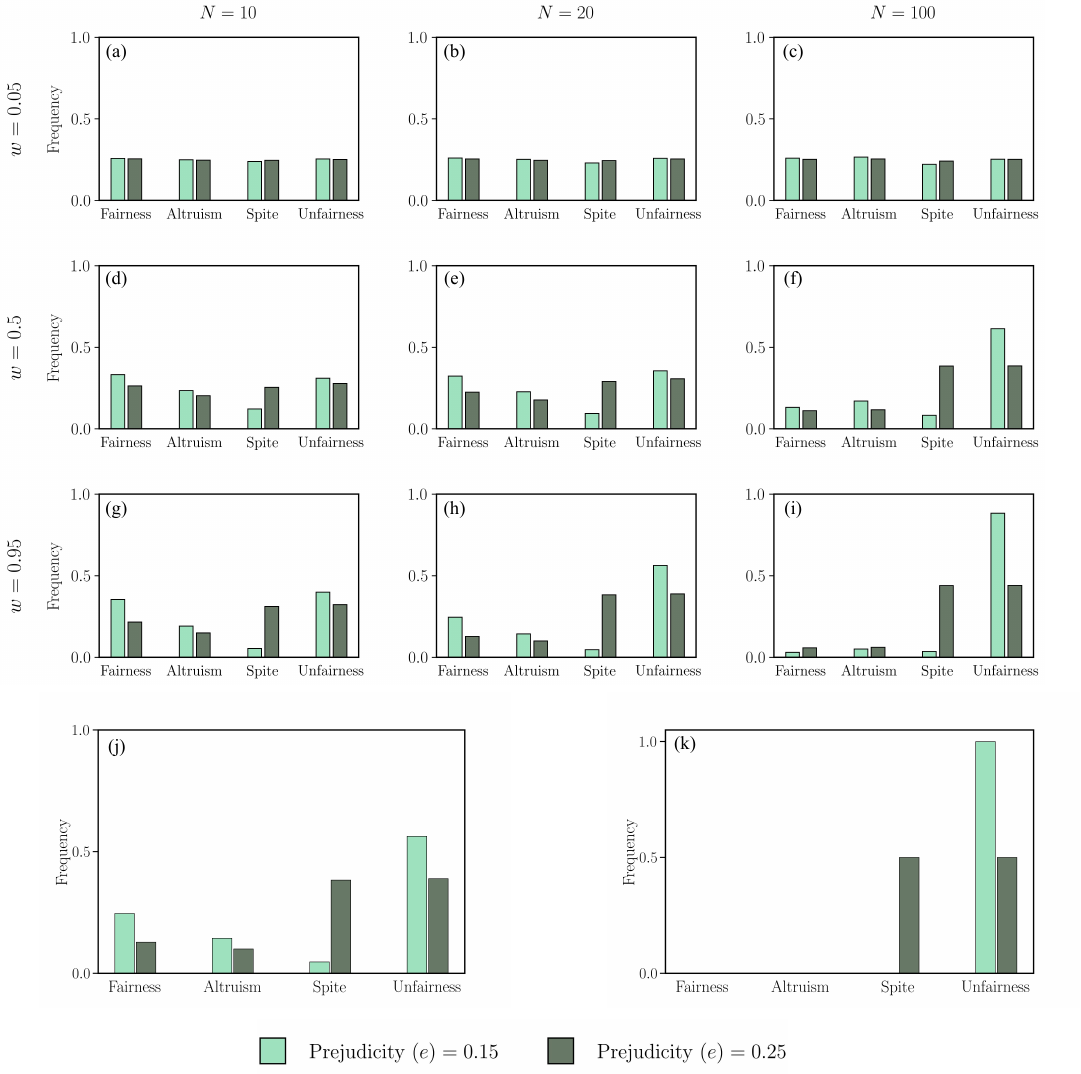}
	\caption {The histogram depicts the frequencies of four strategies: \emph{fairness}, \emph{altruism}, \emph{spite}, and \emph{unfairness}. For each strategy, two bars are shown---light green and dark green---representing cases where prejudicity is below $(e = 0.15 < e_c)$ and above $(e = 0.25 > e_c)$ the critical prejudicity threshold $(e_c=0.2)$, respectively. Subplots (a)--(i), generated from the Moran process, are arranged in a $3 \times 3$ grid corresponding to three selection strengths $w \in \{0.05, 0.5, 0.95\}$ and three population sizes $N \in \{10, 20, 100\}$. Subplot (j) illustrates the results from replicator dynamics in an infinite population, while subplot (k) shows the frequencies of all strategies in the the Imhof--Fudenberg--Nowak process.}
	\label{fig: Spite_vs_prej_all}
\end{figure*}

This section presents a comprehensive investigation of the short-term evolutionary dynamics of all strategies across key parameters---population size, selection strength, mutation rate, and the extent of low offers or demands---and demonstrate that the hallmark discontinuous phase transition remains qualitatively robust across mutation rates. The frequency distribution of four strategies---fairness, altruism, spite, and unfairness---across different population sizes ($N$) and varying selection strengths ($w$) is provided in Fig.~\ref{fig: Spite_vs_prej_all}. Similar to the frequency of spite, the frequency of fairness refers to the total frequency of prejudiced fair and unprejudiced fair strategies. The same applies to the frequencies of unfairness and altruism strategies. In each bar diagram, light green bars represent $e = 0.15$, while dark green bars correspond to $e = 0.25$. For a low selection strength ($w = 0.05$), the frequency of all strategies remains nearly uniform and is largely independent of $e$ and $N$. This is because random drift dominates over the game payoff in determining fitness. Evidently, the frequency of all strategies varies significantly for higher selection strengths ($w = 0.5, 0.95$). For $w = 0.5$, the variation in the frequencies of these strategies becomes conspicuous with increasing population size $N$ under both conditions, $e < e_c$ and $e > e_c$.

When $e < e_c$, the frequency of unfair strategies increases with $N$, while the frequencies of all other strategies decrease. In contrast, when $e > e_c$, both spite and unfair strategies show a nearly equal increase as $N$ grows, while the frequencies of all other strategies decline. Similar trends are observed for the higher selection strength $w = 0.95$. As selection strength increases, the frequency of the unfair strategy becomes significantly higher than all other strategies under the condition $e < e_c$. Meanwhile, under $e > e_c$, the frequencies of spite and unfair strategies increase almost equally with higher selection strength. This is because the payoff in the game contributes more strongly to fitness when selection strength is high. 
\begin{figure*}
	\centering
	\includegraphics[scale=1.11]{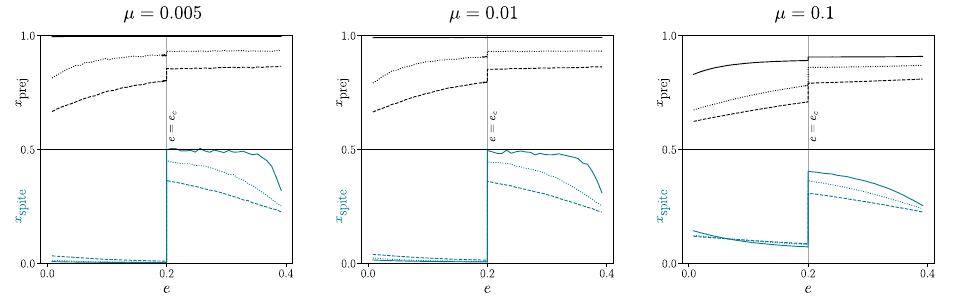}
	\caption {The robustness of the discontinuous phase transition of spite is shown across three mutation rates, $\mu \in \{0.005, 0.01, 0.1\}$. Each subplot shows the frequency of spite ($x_{\text{spite}}$) and the level of prejudice ($x_{\text{prej}}$) as functions of prejudicity for three population sizes $N \in \{10, 20, 100\}$. Teal lines depict the frequency of spite, while black lines represent the level of prejudice. Dashed, dotted, and solid curves correspond to $N = 10$, $N = 20$, and $N = 100$, respectively.}
	\label{fig:phase_transition_over_mutation}
\end{figure*}
\begin{figure*}
	\centering
	\includegraphics[scale=0.8]{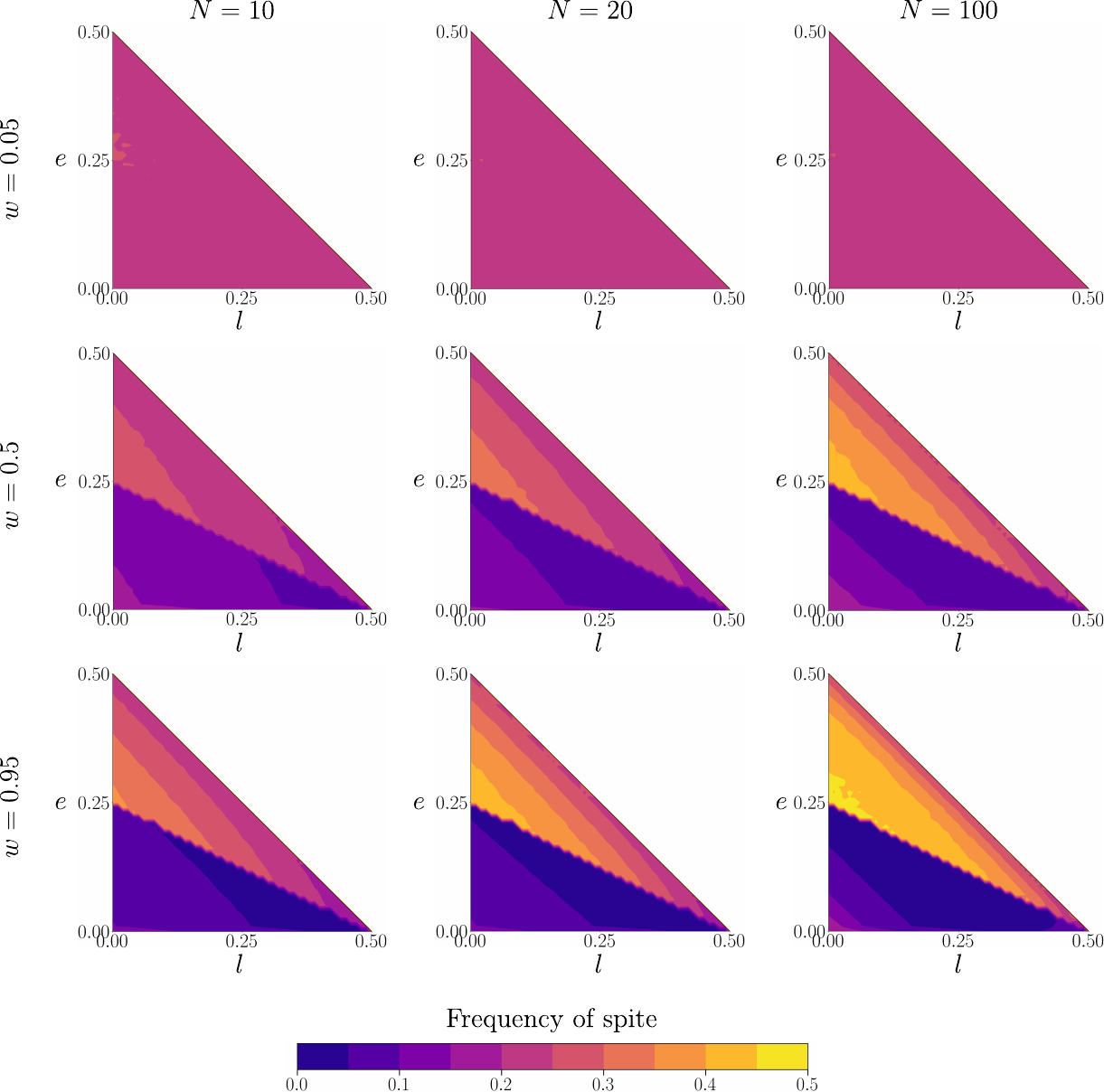}
	\caption{The frequency of spite observed in short-term evolution across the parameter space $e-l$ in a finite population is presented. The subplots, generated using the Moran process, are arranged in a $3 \times 3$ grid. This grid corresponds to three selection strengths: $w \in \{0.05, 0.5, 0.95\}$, and three population sizes: $N \in \{10, 20, 100\}$. The color bar indicates the frequency of spite.}
	\label{fig: Sfreq_contour}
\end{figure*}%
The above analysis highlights that spiteful behavior emerges as the intensity of prejudice crosses the critical value. While the unfair strategy gains prominence with larger population sizes, higher values of $e$ significantly alter the dynamics. Specifically, under the condition $e > e_c$, spiteful and unfair strategies become predominant. This underscores the critical role of prejudice in fostering spiteful behaviors within strategic interactions.

We further examine the emergence of the spiteful strategy through discontinuous phase transitions under different mutation rates to ensure that the parameter values chosen in the main text do not lose qualitative generality. Specifically, we consider mutation rates $\mu = 0.005, 0.01, 0.1$ and find that the discontinuous phase transition remains qualitatively robust across these values; see Fig.~\ref{fig:phase_transition_over_mutation}. The main difference observed is that as the mutation rate increases, the gap in the frequency of spite just below and just above the critical prejudicity value becomes smaller. 

To further rigorously explore the emergence of spite, we present the frequency distribution of the spiteful strategy on a space defined by the prejudicity ($e$) and the minimum offer or demand ($l$), across varying selection strengths ($w$) and population sizes ($N$) in Fig.~\ref{fig: Sfreq_contour}. The contour plots are triangular in shape because $e$ and $l$ follow the condition $e < h - l$. For a low selection strength ($w = 0.05$), the color in the triangular contour plot is almost uniform over $e$ and $l$ for any population size, which suggests that different amounts of low offer or demand do not create any difference in the frequency of spite. This is obvious for the same reason that random drift influences fitness more than the game payoff. For higher selection strengths ($w = 0.5, 0.95$), the frequency distribution varies in the $e$-$l$ space for a given population size. We observe that the frequency of spite decreases as the value of the low offer or demand $l$ increases, for a given prejudicity, whether above or below the critical prejudicity. Taken together, these findings highlight that while prejudice is the dominant factor driving spite, other factors---including the level of low offers or demands, population size, and selection strength---play crucial roles in shaping the emergence of spiteful behaviors.

\section{Short Term Evolution in Infinite Population}\label{sec_app:infinite_population}
As we have observed that the phase transition in larger population is more prominent and robust, it makes sense to study the limit of infinite population and the phase transition may be seen in a mean field model of the evolutionary process. In the frame of evolutionary game theory, we are after a deterministic replication-selection process in the population where the phase transition manifests as changes in the dynamical stabilities of the pure strategies and in sizes of their basins of attraction.

\subsection{Set-up}
We consider an infinite population where $x_{s_i}$ is the frequency of the strategy $s_i\in\mathcal{S}$, i.e., $x_{s_i}\equiv n_{s_i}/N$ in the limit of infinite population. Naturally, the state of the population is described by the frequency profile $(x_{s_1}, x_{s_2} \dots, x_{s_8})$; of course, sum of all the frequencies must be unity. According to the Darwinian tenet, the relative frequency of a fitter strategy increases over time and others gradually die out. 

The evolutionary dynamics of the traits in a well-mixed infinite population is governed by the well known replicator equation~\cite{Schuster1981, Taylor1978, Cressman2014} which encapsulates the key concepts of Darwinian evolution. It describes the change in frequency of a trait proportional to its fitness relative to the average fitness of the population. The equation is
\begin{eqnarray}
	\dot{x}_{s_i}=x_{s_i}(f^w_{s_i} - \bar{f}^w),\label{eq:re}
\end{eqnarray}
where $\bar{f}^w=\sum_{i=1}^{8}x_{s_i}f^w_{s_i}$, is the average fitness of the population. The intensity of selection, $w$, does not change the dynamics except slowing it down---$w$ can be absorbed in time making R.H.S. of Eq.~(\ref{eq:re}) independent of $w$. Hence, in this section, we fix $w=1$ without any loss of generality. 

All the pure eight strategies are fixed points of  Eq.~(\ref{eq:re}). For our purpose, we need not worry about the mixed strategies---interior fixed points---because in the bid to match with the result of Moran process (with $\mu\to0$, $w\to 1$, and $N\to \infty$) only absorbing states of the corresponding Markov chain are of the most interest. Technically, the eigenvalues of the Jacobian matrix about fixed points while carrying out the linear stability analysis plays a crucial role in determining the stability of equilibria in replicator dynamics.
\begin{figure*}[hbt]
	\centering
	\includegraphics[scale=0.765]{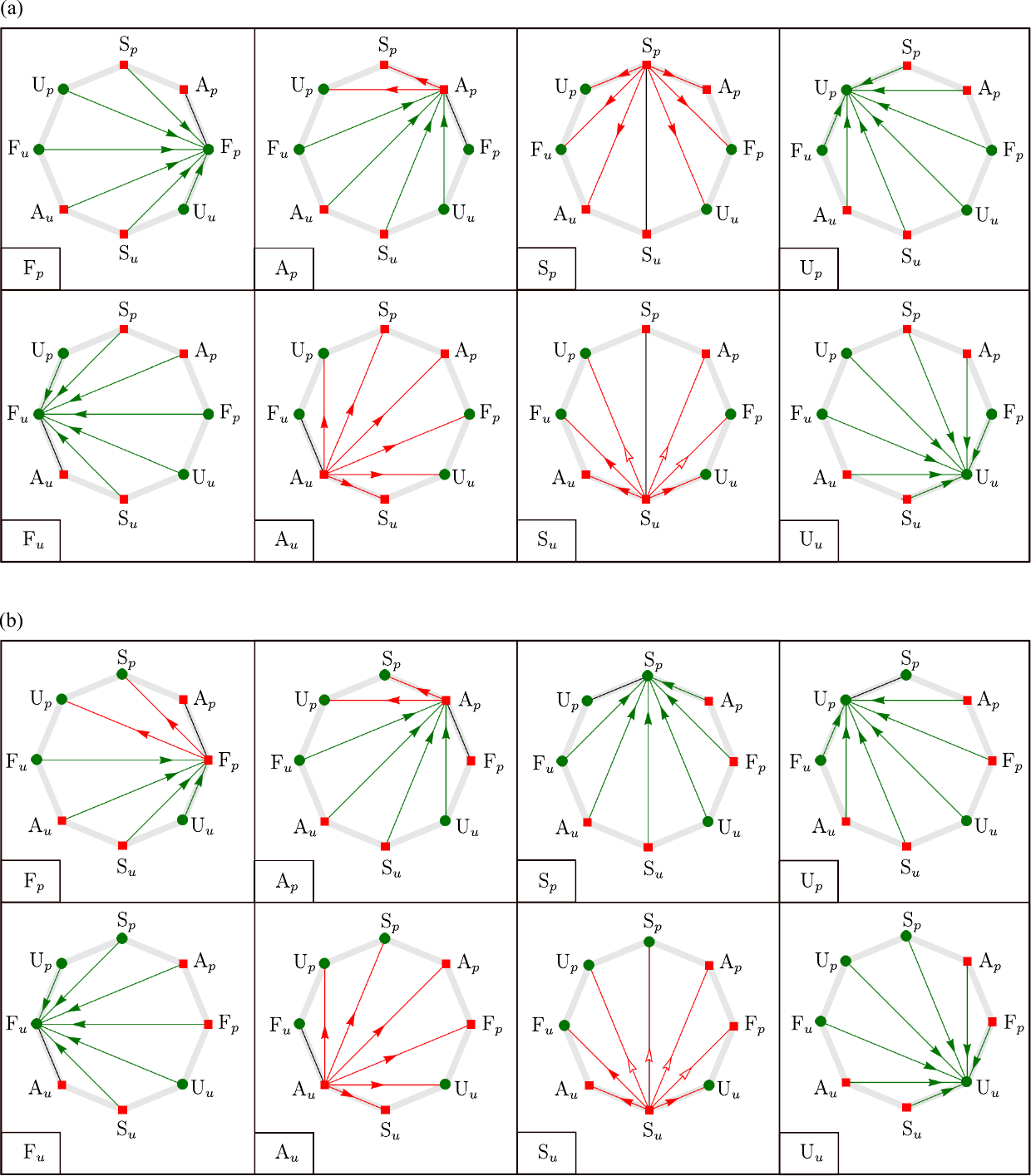}
	\caption {We present the stability diagram of the corner fixed points in a fully connected network of degree seven, where the vertices are the corners of the simplex and the edges of the network correspond to the edges of the simplex. The two subplots show the stability for two different intensities of prejudice: (a) $e<e_c$ and (b) $e_c<e<(h-l)$. The flow is shown for a particular fixed point in each diagram. Flow away from that fixed point is depicted by red arrows, while flow towards the fixed point is depicted by green arrows. The green and red colors indicate stable and unstable fixed points, respectively.}
	\label{fig: stability diagram1}
\end{figure*}
\begin{figure}
	\centering
	\includegraphics[scale=0.30]{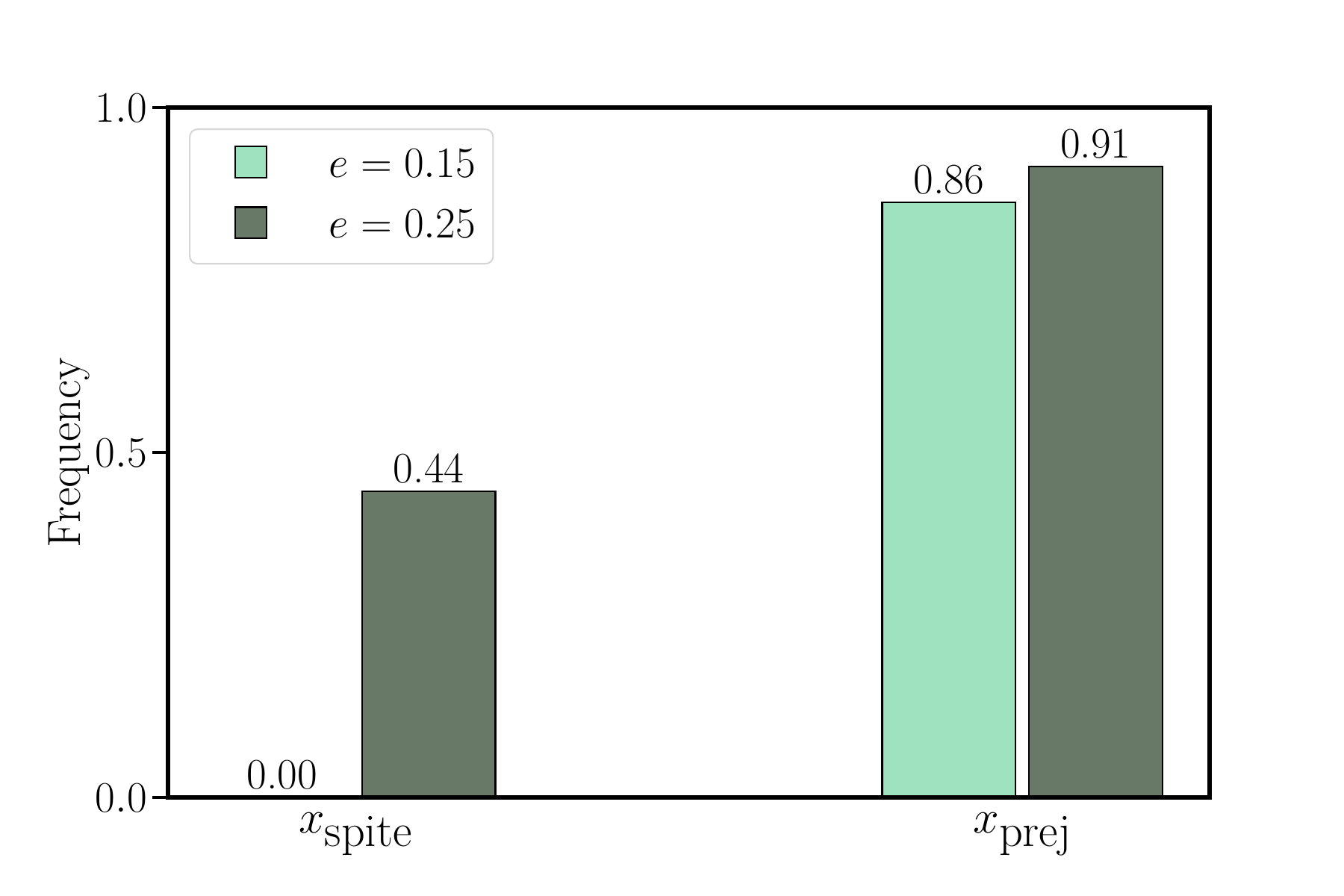}
	\caption {The histogram presents the the frequency of spiteful strategies $(x_{\text{spite}})$ and level of prejudice $(x_{\text{prej}})$ in the short-term evolution of the infinite population. In this context, the frequency of spiteful strategies refers to how many initial conditions reach the fixed points of replicator dynamics at the asymptotic limit, corresponding to both prejudiced and unprejudiced spiteful strategies. The light green and dark green bars, respectively, correspond to prejudicity below $(e = 0.15 < e_c)$ and above $(e = 0.25 > e_c)$ the critical prejudicity ($e_c$). The values of the bars are written on the top of the bars.}
	\label{fig: Spite_vs_prej_inf}
\end{figure}
There is a close connection between dynamic stability of a population state and its being evolutionarily stable states (ESS). 

An ESS~\cite{Smith1973} is a locally asymptotically stable fixed point of the replicator equation~\cite{Zeeman1980, Hofbauer1979, Bomze1991} and a weak ESS~\cite{Nowak2006Book} is a locally Lyapunov stable fixed point of the replicator equation. Staying within the collection of pure strategies, suppose there is a strategy $\hat{s}\in \mathcal{S}$ such that (i) $\pi(\hat{s},\hat{s})\ge \pi(s_i,\hat{s})$ $\forall s_i$, (ii) and if $\pi(\hat{s},\hat{s})\ge \pi(s_i,\hat{s})$ for some $s_i=s_j$, then  $\pi(\hat{s},s_j)\ge \pi(s_j,s_j)$. Then $\hat{s}$ is a weak ESS, and it is an ESS if the inequality in the last condition is a strict inequality. 

In the context of this paper, $\pi(s_i,s_j)$ (the payoff corresponding to strategy $s_i$ against $s_j$) can be easily read from the payoff matrices presented in Table~\ref{tab_1}. Consequently, these game-theoretic equilibria can be directly calculated from the payoff matrices, as done for our system in Table~\ref{tab:rps_ess}. As an illustration, consider the strategy $U_p$, which is an ESS when $0<e<e_c$ and a weak ESS when $e_c\leq e<h-l$. First, take the case $0<e<e_c$. The condition (i), $\pi(U_p,U_p)\geq \pi(s_i,U_p)$, must hold for all $s_i\in\mathcal{S}$ for $U_p$ to be an ESS. Suppose $s_i=A_p$; then the condition $1 \geq \pi(A_p,U_p)$ needs to be satisfied since $\pi(U_p,U_p)=1$. This holds whenever $e\leq h-l$, which is satisfied because $e<\tfrac{h-l}{2}$. Checking similarly for all other $s_i$ shows that $\pi(U_p,U_p)>\pi(s_i,U_p)$ holds for every $s_i\in\mathcal{S}$. Hence, $U_p$ is an ESS. Next, consider the case $e_c\leq e<h-l$. Again, $\pi(U_p,U_p)=1$, and it appears that $\pi(U_p,U_p)=\pi(s_i,U_p)$ for $s_i=S_p$. We then check the condition (ii), $\pi(U_p,S_p)\geq \pi(S_p,S_p)$, and find that it holds. Thus, $U_p$ is a weak ESS for $e_c\leq e<h-l$.

\subsection{Results}

Comprehensive linear stability analysis about the pure strategy fixed points of the replicator equation is diagrammatically depicted in the stability diagrams in Fig.~\ref{fig: stability diagram1}. Each subfigure represents the stability of a specific strategy, labeled at its lower-left corner. The four red squares and the four green circles (seemingly lying on vertices of an octagon) correspond to the eight distinct strategies of set $\mathcal{S}$. 

A filled green arrowhead along a diagonal/edge pointing toward a green circle (a fixed point) indicates its stability, characterized by a negative real part of eigenvalue (of the Jacobian of the linearized dynamics about the fixed point),  along that diagonal/edge (a stable eigen-direction). Similarly, a filled red arrowhead along a diagonal/edge pointing away a red square indicates its instability, characterized by a positive real part of eigenvalue,  along that diagonal/edge (an unstable eigen-direction). Finally, rest of the diagonals/edges emanating from a vertex (either a square or a circle) denote neutral stability along that direction within the paradigm of linear stability; this is characterized by eigenvalue with zero real part. However, since an eigenvalue with zero real part basically indicates that the linear stability has failed, we resort to numerics to determine the true nature of stability. We introduce an initial condition on the corresponding eigen-direction in the neighbourhood of the fixed point and observe its future. The conclusions, thus found, are depicted in the Figure in the following way: The diagonal/edge has no arrowhead if even numerics (i.e., going beyond linear stability) suggests neutral stability (this is indicative of non-isolated fixed pointed which we have verified). Otherwise,  the diagonal/edge has either an open red or an open green arrowhead indicating locally repelling or locally attracting nature of the fixed point, respectively. Locally asymptotically stable fixed points are the ones which possess only eigen-directions with either open or filled green arrowhead, while locally Lyapunov stable fixed points are the ones with at least one eigen-direction with no arrowhead and no eigen-direction with either open or filled red arrowhead; any other fixed point is unstable. 
\begin{table}
	\centering
	\begin{tabular}{|c|c|c|}
		\hline
		&    \multicolumn{2}{c|}{ }\\
		\multirow{3}{*}{Strategy}  &    \multicolumn{2}{c|}{Evolutionary Stability}\\		
		
		\cline{2-3}
		&&\\
		& $0<e<e_c$            & $e_c\le e < h-l$  \\ 
		
		\hline
		
		$\text{F}_p= (\Theta_p, \text{H}, \text{H})$ &   weak ESS        & not ESS  \\ 
		
		\hline
		
		$\text{A}_p= (\Theta_p, \text{H}, \text{L})$&  not ESS          & not ESS  \\
		\hline

		$\text{S}_p= (\Theta_p, \text{L}, \text{H})$&   not ESS          & weak ESS  \\
		\hline

		$\text{U}_p= (\Theta_p, \text{L}, \text{L})$&   ESS         & weak ESS \\
		\hline

		$\text{F}_u= (\Theta_u, \text{H}, \text{H})$&   weak ESS         & weak ESS  \\ 
		\hline

		$\text{A}_u= (\Theta_u, \text{H}, \text{L})$&   not ESS          & not ESS   \\
		\hline

		$\text{S}_u= (\Theta_u, \text{L}, \text{H})$&  not ESS         & not ESS   \\
		\hline
		
		
		$\text{U}_u= (\Theta_u, \text{L}, \text{L})$&   ESS        & ESS    \\
		\hline
		
	\end{tabular}
	\caption{Evolutionarily Stable Strategy (ESS) in the  game.}
	\label{tab:rps_ess}
\end{table}
In Fig.~\ref{fig: stability diagram1}(a), stability for \(e < e_c\) has been shown. In this case, the fixed points corresponding to fair and unfair strategies (\(\text{F}_p , \text{F}_u, \text{U}_p, \text{U}_u\)) exhibit stability: Specifically, \(\text{F}_p\) and \(\text{F}_u\) are Lyapunov stable, whereas \(\text{U}_p\) and \(\text{U}_u\) are asymptotically stable. It is assuring that the former pair are weak ESSes and the latter ones are ESS as per Table~\ref{tab:rps_ess}. Rest of the four strategies are not stable and hence not ESS as well.
\begin{figure*}
	\centering
	\includegraphics[scale=0.9]{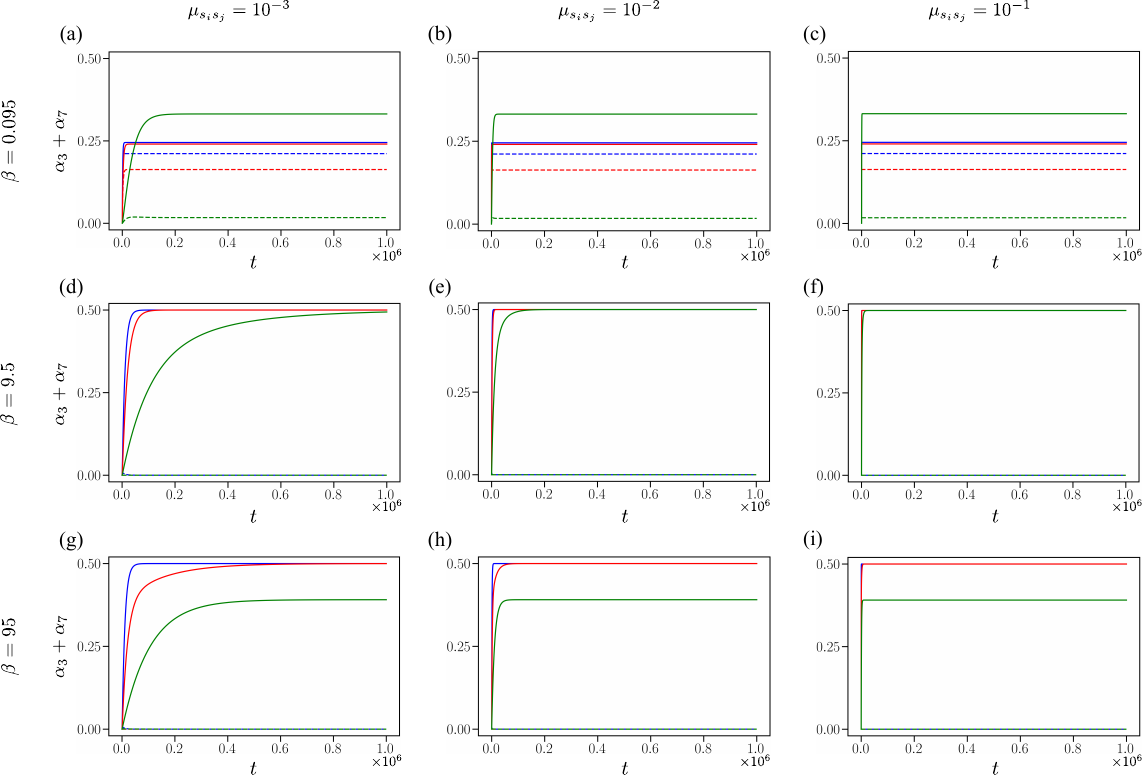}
	\caption{Time evolution of the frequency of spiteful strategies below and above the critical prejudicity for selection strengths $\beta \in \{0.095, 9.5, 95\}$ and mutation rates $\mu_{s_is_j} \in \{10^{-3}, 10^{-2}, 10^{-1}\}$ in the Imhof--Nowak--Fudenberg process. The frequency of spite denotes the number of occurrences of a spiteful monomorphic population up to time $t$. Dashed and solid lines correspond to prejudicity values $e=0.15<e_c$ and $e=0.25>e_c$, respectively. Blue, red, and green curves indicate population sizes $N=10$, $N=20$, and $N=100$, respectively.}
	\label{fig:selection_strength_vs_mutaion_rate}
\end{figure*}
Likewise, in Fig.~\ref{fig: stability diagram1}(b), stability for \(e > e_c\) has been presented. Beyond the critical threshold prejudicity \(e_c\), a different set of ESSes appear. The prejudiced unfair strategy (\(\text{U}_p\)) remians stable but with weakened stability: It becomes Lyapunov stable only. The prejudiced fair strategy (\(\text{F}_p\)) loses stability to become an unstable strategy reflecting its diminished evolutionary viability. The spiteful strategy (\(\text{S}_p\)), on the other hand,  gains stability, emerging as a Lyapunov stable fixed point. This shift highlights the evolutionary advantage of spiteful behaviour as prejudicity increases. $F_u$ and $U_u$ do not change their types---they remain, respectively, Lyapunov and asymptotically stable; and hence, weak ESS and ESS, respectively. Other unstable strategies remain unstable in this case as well. The expected connection between game-theoretic equilibria (see Table~\ref{tab:rps_ess}) and dynamic equilibria obviously holds good.

In summary, the eigenvalue-driven stability shifts have directly linked to the observed phase transition as prejudice has crossed the critical threshold \(e_c\). Below \(e_c\), the system has stabilized around fair and unfair strategies. Above \(e_c\), spiteful behaviours have gained stability. The results of the eigenvalue-driven stability analysis closely match the outcomes of the static ESS analysis, with both approaches identifying same stability patterns and transitions. Together, these analyses provide a comprehensive view of how prejudice influences the evolutionary dynamics of social interactions, highlighting key transitions in stability and the emergence of spiteful behaviours.

Next, we go beyond the domain of linear stability analysis. Specifically, for obvious reasons, we expect the subplot corresponding to $w=0.95\approx 1$ and $N=100$ (large $N$) in Fig.~\ref{fig: Spite_vs_prej} to emerge in the case of replicator dynamics as well. To this end, to analyze the evolutionary fate of spiteful strategy and level of prejudice in infinite population, the replicator dynamics have been numerically solved over an extended period, starting from a uniformly distributed initial condition on the phase space---partitioned into a grid of step size \(0.05\)---and result is presented in Fig.~\ref{fig: Spite_vs_prej_inf}.  The analysis has been restricted to pure fixed points.  
(In passing, we mention that \(75\%\) of the trajectories converged to pure fixed points when \(e < e_c\), whereas \(42\%\)  converged to these points when \(e \ge e_c\); rest of the trajectories converge to the interior fixed points which have not been considered in our study.) The frequency distribution of a specific strategy (say, \(s_i\)) has been calculated as the ratio of trajectories converging to \(s_i\) to the total number of trajectories converging to all pure fixed points. The resulting frequency distribution of spiteful strategies in infinite populations has been plotted in Fig.~\ref{fig: Spite_vs_prej_inf}. It has been shown that when \(e < e_c\), the frequency of spiteful strategies has remained nearly zero. However, when \(e \ge e_c\), the frequency of spiteful strategies has emerged significantly. These findings closely align with results obtained for finite populations under high selection strength and a large population size---compare Fig.~\ref{fig: Spite_vs_prej}(i) with Fig.~\ref{fig: Spite_vs_prej_inf}. We refer the interested readers to Fig.~\ref{fig: Spite_vs_prej_all}(j) for inspecting the frequency distribution of other strategies in infinite population.
\section{Long-term Evolution in Finite Population} \label{sec_app:long_term_finite_population}

Here we provide an account of the long-term evolutionary dynamics of spite across different mutation rates and selection strengths. The frequency distribution of the four strategies—fairness, spite, altruism, and unfairness—for selection strength $\beta=0.95$ and mutation rate $\mu=10^{-3}$ is also shown in Fig.\ref{fig: Spite_vs_prej_all}(k) using a bar diagram, with colors carrying the same meaning as in earlier cases. When prejudicity is below the critical value, only the unfair strategy survives in the long term, as indicated by the light green bar in Fig.\ref{fig: Spite_vs_prej_all}(k). Similar to the short-term evolutionary scenario, the frequency of spite rises sharply once prejudicity crosses a critical threshold. Beyond this point, both spite and unfairness persist, co-evolving in the long-term dynamics.

An investigation of spite across different mutation rates and selection strengths shows that varying mutation rates does not significantly alter the outcomes. For selection strength $\beta<1$, the results remain qualitatively similar (see Fig.~\ref{fig: SFreq_timej} and Figs.~\ref{fig:selection_strength_vs_mutaion_rate}(a)–(c)). However, the outcomes change qualitatively when selection is sufficiently strong and prejudicity lies above the critical value. For instance, when $\beta=95$, the green solid line representing the abundance of spite for population size $N=100$ lies below the blue and red solid lines corresponding to $N=10$ and $N=20$, respectively (see Figs.~\ref{fig:selection_strength_vs_mutaion_rate}(g)-(i)). This indicates that the abundance of spite decreases as population size increases under strong selection. It holds for different mutation rates as well (see Figs.~\ref{fig:selection_strength_vs_mutaion_rate}(d)-(f) and see Figs.~\ref{fig:selection_strength_vs_mutaion_rate}(g)-(i)). This decline in the prevalence of spite with larger populations can be attributed to the fact that spite is only a weak ESS in the infinite population limit (see Table~\ref{tab:rps_ess}). Consequently, spiteful residents are more easily invaded by mutants in larger populations, leading to a reduction in the long-term frequency of spite.

\bibliography{patra_etal_reference.bib}
	
\end{document}